\pdfoutput=1
\documentclass[final,authoryear,5p,times,twocolumn]{elsarticle}

\usepackage{lineno}
\usepackage{amsmath,amssymb}
\usepackage[colorlinks=True,
            urlcolor=blue,
            linkcolor=blue,
            citecolor= blue,
	    pdftex]{hyperref}
\usepackage{journals} 
\usepackage{booktabs}

\journal{Icarus}

\newcommand{\lp}{\left(}
\newcommand{\rp}{\right)}
\newcommand{\tb}{\bar{T}}
\newcommand{\rb}{\bar{\rho}}

\newcommand{\vna}{\vec{\nabla}}

\def\vec#1{\ensuremath{\mathchoice{\mbox{\boldmath$\displaystyle#1$}}
{\mbox{\boldmath$\textstyle#1$}}
{\mbox{\boldmath$\scriptstyle#1$}}
{\mbox{\boldmath$\scriptscriptstyle#1$}}}}
\def\tens#1{\ensuremath{\mathsf{#1}}}

\begin{document}

\begin{frontmatter}

\title{Effects of compressibility on driving zonal flow in gas giants}

\author[MPS]{T.~Gastine\corref{cor1}}
\ead{gastine@mps.mpg.de}
\author[MPS]{J.~Wicht}

\address[MPS]{Max Planck Institut f\"ur Sonnensytemforschung, Max Planck Strasse
2, 37191 Katlenburg Lindau, Germany}

\cortext[cor1]{Principal corresponding author}

\begin{abstract}
The banded structures observed on the surfaces of the gas giants are associated
with strong zonal winds alternating in direction with latitude.
We use three-dimensional numerical simulations of
compressible convection in the anelastic approximation to explore
the properties of zonal winds in rapidly rotating spherical shells.
Since the model is restricted to the electrically insulating outer envelope,
we therefore neglect magnetic effects.

A systematic parametric study for various density scaleheights and Rayleigh
numbers allows to explore the dependence of convection and zonal jets on these
parameters and to derive scaling laws.

While the density stratification affects the local flow amplitude and the
convective scales, global quantities and zonal jets properties remain fairly
independent of the density stratification.
The zonal jets are maintained by Reynolds stresses, which rely on the
correlation between zonal and cylindrically radial flow components. The gradual
loss of this correlation with increasing supercriticality hampers all our
simulations and explains why the additional compressional source of vorticity
hardly affects zonal flows.

All these common features may explain why previous Boussinesq models were
already successful in reproducing the morphology of zonal jets in gas giants.
\end{abstract}


\end{frontmatter}

\section{Introduction}

The banded structures at the surfaces of Jupiter and Saturn are associated with
prograde and retrograde zonal flows. In both planets, a large
amplitude eastward equatorial jet (150 $\text{m}.\text{s}^{-1}$ for Jupiter
and 300 $\text{m}.\text{s}^{-1}$ for Saturn) is flanked by multiple alternating
zonal winds of weaker amplitudes (typically tens of $\text{m}.\text{s}^{-1}$).
This alternating pattern is observed up to the polar region \citep{Porco03}.

Two competing types of models have tried to address the question of
what drives the zonal winds and how deep they reach into the planet
\citep[see the review of][]{Vasavada05}.
In the ``weather layer'' hypothesis, zonal winds are confined
to a very thin layer near the cloud level. Such shallow models, that rely
on global circulation codes, successfully recover the observed alternating
banded structures of zonal flows \citep[e.g.][]{Williams78,Cho96}. While
earlier shallow models mostly produce rather retro- than prograde equatorial 
jet, this
has been hampered in most recent approaches by adding
additional forcing mechanisms,
such as moist radiative relaxation at the equator \citep{Scott08} or
condensation of water vapor \citep{Lian10}. In the ``deep models'',
the zonal winds are supposed to extend over the whole molecular envelope ($\sim
10^{4}\text{~km}$). The direction, the amplitude and the number of jets are
reproduced, provided thin shells are assumed and convection is strongly driven
\citep{Heimpel05}.

The \textit{in-situ} measurement by the Galileo probe further stimulated the
discussion. They showed that the velocity increases up to
170~$\text{m}.\text{s}^{-1}$ over the sampling depth between $0.4$ and 22 bars
 \citep{Atkinson97}. Though this result indicates
that the amplitude of zonal winds increases well below the cloud level,
it hardly proves the ``deep models'' since the probe has merely scratched the
outermost 150~kms of Jupiter's atmosphere \citep{Vasavada05}.

Interior models of the giants planets suggest that Hydrogen becomes metallic at
about 1.5~Mbar, corresponding roughly to 0.85 Jupiter's radius and 0.6 Saturn's
radius \citep{Guillot99,Guillot04,Nettelmann08}. This has traditionally been
used to independently model the dynamics of the molecular layer without
regarding the conducting region. Lorentz forces and increasing density would
prevent these strong zonal winds to penetrate significantly into the molecular
layer, where timescales thus tend to be slower. However,
shock waves experiments suggest that the increase of electrical
conductivity is more gradual rather than abrupt \citep{Nellis96,Nellis00}.
\cite{Liu08} therefore argue that zonal jets must be confined to a thin upper
layer (0.96 Jupiter's radius and 0.86 Saturn's radius), since deep zonal
winds would already generate important Ohmic
dissipation incompatible with the observed surface flux.
However, \cite{Glatz08} claimed that the kinetic model of \cite{Liu08} is too
simplistic as this does not allow the magnetic field to adjust the differential
rotation, which would severely reduce Ohmic dissipation.

Notwithstanding this open discussion, the fact that deep models successfully
reproduce the primary properties of the zonal jets speaks in their favour.
These models rely on 3-D numerical simulations of rapidly rotating shells
\citep[e.g.][]{Christensen01,Christensen02,Heimpel05}. When convection is
strongly driven, the nonlinear inertial term $\vec{u}\cdot\vec{\nabla}\vec{u}$
becomes influential and gives rises to Reynolds stresses, a statistical
correlation between the convective flow components that allows to feed energy
into zonal winds \citep[e.g.][]{Plaut08}.

Most of these simulations use the Boussinesq approximation, which assumes
homogeneous reference state. This seems acceptable for terrestrial planets,
but becomes rather dubious for gas planets, where for example the density
increases by four orders of magnitude \citep{Guillot99,Anufriev05}.

The anelastic approximation we adopt here, provides
a more realistic framework, which allows to incorporate the background density
stratification, while filtering out fast acoustic
waves \citep{Lantz99}. \cite{Jones09} presented 3-D anelastic simulations that
showed many interesting differences with the Boussinesq results.
The zonal flows still remain large-scale and deep-seated but the local
small-scale convection is severely affected by compressibility and strongly
depends on radius. While the main equatorial band remains a robust
common feature, it seems more difficult to get stable
multiple high latitude jets, even in the most extreme parameters.  
\cite{Kaspi09} used a modified anelastic approach neglecting
viscous heating but including a more realistic equation of state. In their
model, the zonal flows show a more pronounced variation in the direction of the
rotation axis than in the work of \cite{Jones09}.
\cite{Evonuk08} and \cite{Glatz09} pointed out that
compressibility adds a new vorticity source that could potentially help
to generate Reynolds stresses that, \textit{in fine}, maintain zonal flows.

Following theses recent studies, we focus here on the effects of
compressibility on rapidly rotating convection. The main aim of this paper is
to determine the exact influence of the density background on the driving
of zonal flows.  To this end, we have conducted a systematic parametric study
for various density stratifications and solutions that span weakly
supercritical to strongly nonlinear convection.

In section~\ref{sec:model}, we introduce the anelastic model and the numerical
methods. Section~\ref{sec:flowprop} presents the results, starting with
convection close to the onset and then progressing into the nonlinear regime. In
section~\ref{sec:scaling}, we concentrate on the zonal winds mechanism and
related scaling laws, before concluding in section~\ref{sec:conclusion}.

\section{The hydrodynamical model}

\label{sec:model}

\subsection{The anelastic equations}

We consider thermal convection of an ideal gas in a spherical shell of outer
radius $r_o$ and inner radius $r_i$, rotating at a constant frequency $\Omega$
about the $z$ axis. Being interested in the dynamics of the molecular region of
gas giants, we assume that the mass is concentrated in the inner part, resulting
in a gravity $g\propto 1/r^2$. We employ the anelastic approximation
following \cite{Brag95}, \cite{Glatz1} and \cite{Lantz99}.
Thermodynamical quantities, such as density, temperature and pressure are
decomposed into the sum of an adiabatic reference state (quantities with
overbars) and a perturbation (primed quantities):
\begin{equation}
 \rho = \rb + \rho' \quad;\quad T = \tb+T' \quad;\quad p = \bar{p}+p'.
\end{equation}
When assuming $g\propto 1/r^2$, the polytropic and adiabatic reference
state is given by \citep[see][for further details]{Jones11}

\begin{equation}
  \tb(r) = \dfrac{c_0}{(1-\eta)r}+ 1-c_0 \quad\text{and}\quad\rb(r) = \tb^m,
\end{equation}
with

\begin{equation}
c_0 = \dfrac{\eta}{1-\eta}\lp \exp\dfrac{N_\rho}{m} -1 \rp \quad\text{with}\quad
N_\rho = \ln\dfrac{\rb(r_i)}{\rb(r_o)}.
\end{equation}
Here $\tb$ and $\rb$  are the background temperature and density,
$m$ is the polytropic index, $\eta=r_i/r_o$ is the aspect ratio and $N_\rho$
corresponds to the number of density scale heights covered over the layer.
For example, $N_\rho=3$ corresponds to a density contrast of approximately 20,
while the largest value explored here ($N_\rho=5$) implies $\rb_i/\rb_o \simeq
150$.

We use a dimensionless formulation, where outer boundaries reference values
serve to non-dimensionalise density and temperature. The shell thickness
$d=r_o-r_i$ is used as a length scale, while the viscous diffusion time
$d^2/\nu$ serves as the time scale, $\nu$ being the constant kinematic
viscosity. Entropy is expressed in units of $\Delta s$, the small imposed
entropy contrast over the layer. The anelastic continuity equation is

\begin{equation}
 \vec{\nabla}\cdot \lp \rb \vec{u} \rp = 0,
 \label{eq:anel}
\end{equation}
while the dimensionless momentum equation is then

\begin{equation}
   \text{E}\lp\dfrac{\partial \vec{u}}{\partial
   t}+\vec{u}\cdot\vec{\nabla}\vec{u}\rp
   +2\vec{e_z}\times\vec{u}=
 -\vec{\nabla}{\dfrac{p'}{\rb}}+\dfrac{\text{Ra}\,\text{E}}{\text{Pr}
}\dfrac{r_o^2}{r^2}s\,\vec{e_r} +
 \dfrac{\text{E}}{\rb} \vec{\nabla}\cdot\tens{S},
 \label{eq:NS}
\end{equation}
where $\vec{u}$, $p$ and $s$ are velocity, pressure and entropy, respectively.
$\tens{S}$ is the traceless rate-of-strain tensor with a constant kinematic
viscosity, given by

\begin{equation}
\tens{S}=\rb\left(\frac{\partial u_i}{\partial x_j} +
\frac{\partial u_j}{\partial x_i}-\frac{2}{3} \delta_{ij}
\vec{\nabla}\cdot\vec{u}\right).
\label{eq:tenseur}
\end{equation}
The dimensionless entropy equation then reads

\begin{equation}
\rb\tb\lp\dfrac{\partial s}{\partial t} + \vec{u}\cdot\vec{\nabla} s\rp =
\dfrac{1}{\text{Pr}}\vec{\nabla}\cdot\lp\rb\tb \vec{\nabla} s\rp +
\text{Di}\,\rb\tens{S}^2,
\label{eq:entropy}
\end{equation}
where thermal diffusivity is assumed to be constant. As stated  by
\cite{Jones09}, in anelastic simulations with large density stratification,
viscous heating plays a significant role in the global energy balance. It
involves the dissipation parameter Di \citep[e.g.][]{Anufriev05}, that is given
by

\begin{equation}
 \text{Di} = \dfrac{\eta\text{Pr}\lp e^{N_\rho/m} -1\rp}{\text{Ra}}.
\end{equation}
In this formulation, turbulent viscosity is expected to dominate over
molecular viscosity and therefore thermal diffusion relies mainly on entropy
diffusion rather than on temperature diffusion
\citep{Brag95,Lantz99,Jones09,Jones11}. That is why, the turbulent heat flux has
been assumed to be proportional to the entropy gradient as in the
mixing-length theory \citep{BV1,BV2}.

\begin{figure}
 \centering
 \includegraphics[width=8cm]{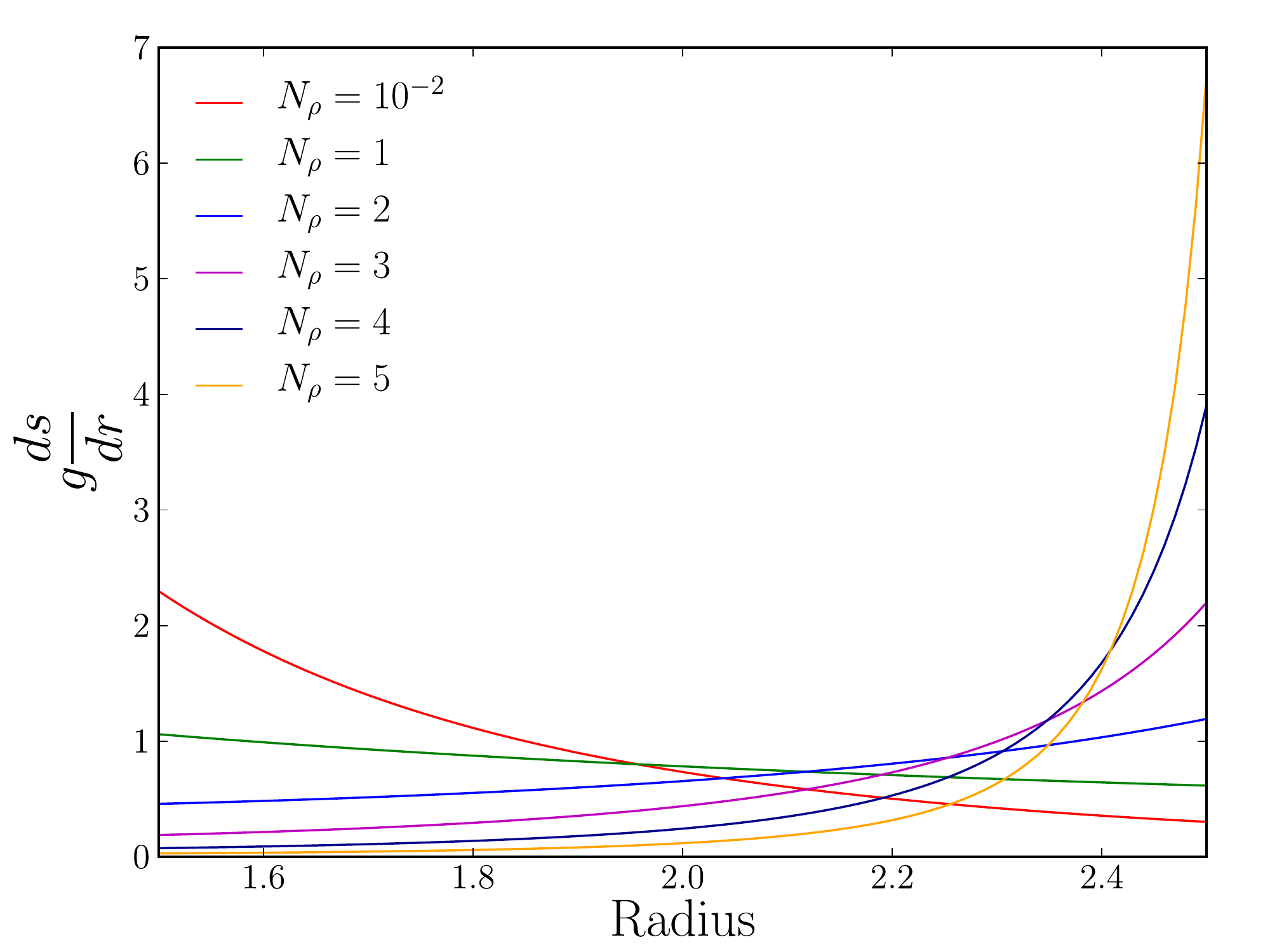}
 \caption{Radial profile of $g |ds_c/dr|$ for different background density
stratifications $N_\rho = [10^{-2}, 1, 2, 3, 4, 5]$.}
\label{fig:raprof}
\end{figure}

In addition to the two anelastic parameters (the polytropic index $m$ and the
density stratification $N_\rho$) and the aspect ratio $\eta$, the three
non-dimensional parameters that control the system of equations (\ref{eq:anel}),
(\ref{eq:NS}) and (\ref{eq:entropy}) are the Ekman number

\begin{equation}
 \text{E} = \dfrac{\nu}{\Omega d^2},
\end{equation}
the Prandtl number

\begin{equation}
 \text{Pr} =
\dfrac{\nu}{\kappa},
\end{equation}
and the Rayleigh number

\begin{equation}
 \text{Ra} = \dfrac{g_0 d^3 \Delta s}{c_p \nu\kappa},
 \label{eq:racode}
\end{equation}
$g_0$ being the gravity at the outer radius, $c_p$ the specific heat, and
$\kappa$ the thermal diffusivity. This definition of the Rayleigh
number is based on the entropy jump over the layer and on values of the physical
quantities at the outer boundary. However, for analysing where
convection sets in first, the radial dependence of the diffusive entropy
gradient has to be considered. This can be cast into a depth-dependent Rayleigh
number:

\begin{equation}
{\cal R} = \dfrac{ g
d^4}{c_p\nu\kappa}\left|\dfrac{ds_c}{dr}\right|,
\label{eq:raloc}
\end{equation}
where $s_c$ is the solution of \citep[see][]{Jones11}
\begin{equation}
 \vec{\nabla}\cdot\lp\rb\tb \vec{\nabla}s_c\rp = 0.
\label{eq:entropyback}
\end{equation}
In our simulations, thermal and viscous diffusivities are assumed to be
constant, but gravity and entropy vary with radius. Gravity is decreasing 
outward (as $g\propto 1/r^2$), but the entropy
gradient is inversely proportional to $\rb\ \tb$
(Eq.~\ref{eq:entropyback}) and therefore increases rapidly toward the surface 
for large density stratifications. As a consequence, we
can see on Fig.~\ref{fig:raprof} that this local Rayleigh number increases
outward for density stratification larger than 3, while it increases downward
for weaker density stratifications.

Several authors argued that Rayleigh numbers at mid-depth may offer a more
meaningful reference value  \citep[e.g.][]{Unno60,Gough76,Glatz2}. Here we
follow \cite{Kaspi09}, who suggest to use a mass-weighted average defined as

\begin{equation}
 \langle{\cal R} \rangle_\rho =  \dfrac{d^4}{c_p \nu\kappa}\left\langle g
\left|\dfrac{ds_c}{dr}\right|\right\rangle_\rho,
\label{eq:ramass}
\end{equation}
where the average is defined as $\langle\cdots\rangle_\rho =
\int_{r_i}^{r_o} (\cdots) \rb r^2 dr /\int_{r_i}^{r_o} \rb r^2 dr$
\citep{Kaspi09}. This new definition of the Rayleigh number can be related to
the usual one (Eq.~\ref{eq:racode}), by the coefficient $\langle g |ds/dr|_c
\rangle_\rho$, that has an analytical solution for a polytropic index of $2$

\begin{equation}
 \left\langle g\left|\dfrac{ds_c}{dr}\right| \right\rangle_\rho =
\dfrac{3N_\rho}{\lp 1-e^{-N_\rho}\rp\lp 1+\eta e^{N_\rho/2}+\eta^2
e^{N_\rho}\rp}.
\label{eq:rescale_anel}
\end{equation}
Moreover, in the limit of small viscosity and thermal diffusivity, the force
balance is expected to become independent on viscosity (low Ekman number).
Therefore, following \cite{Christensen02} and \cite{Christensen06}, it is
more relevant to consider a modified Rayleigh number that does not depend
anymore on diffusivities, defined as

\begin{equation}
 \left\langle\text{Ra}^*\right\rangle_\rho
= \dfrac{1}{c_p \Omega^2}\left\langle g\left|\dfrac{ds_c}{dr}\right|
\right\rangle_\rho=
\langle{\cal R} \rangle_\rho\,\text{E}^2\,\text{Pr}^{-1}.
\end{equation}
To get a non-dimensional measure of the radial heat transport that is also
independent of the thermal diffusivity, we use a modified Nusselt number

\begin{equation}
 \text{Nu}^* =\dfrac{q}{\rho T \Omega d^2 (ds_c/dr)} =
\text{Nu}\,\text{E}\,\text{Pr}^{-1}.
\end{equation}
where $q$ is the heat flux. Once the simulation has reached the nonlinear
saturation, the time-average of this modified Nusselt number should become
constant with radius as there is no internal heat source in the system.
Therefore, there is no need to consider mass-weighted average of this
parameter.

Finally, it is also useful to analyse our simulations in terms of an alternate
Rayleigh number based on the heat flux rather than on the entropy contrast
$\Delta s$

\begin{equation}
 \hbox{Ra}_q^* = \dfrac{q g}{c_p \rho T \Omega^3 d^2}.
\end{equation}
This flux-based Rayleigh number can be related to the modified Nusselt number as
follows

\begin{equation}
 \hbox{Ra}_q^* = \dfrac{g}{c_p \Omega^2}\left|\dfrac{ds_c}{dr}\right|
\text{Nu}^* =
\text{Ra}^*\,\text{Nu}^*.
\label{eq:fluxRa}
\end{equation}
Once again, this number varies with radius, and therefore it is
useful to consider the mass-weighted average counterpart of this quantity

\begin{equation}
 \langle\hbox{Ra}_q^*\rangle_\rho =
\left\langle\hbox{Ra}^*\right\rangle_\rho\,\text{Nu}^*.
\label{eq:fluxRarho}
\end{equation}

\subsection{The numerical method}

The numerical simulations of this study have been carried out with a modified
version of the code MagIC \citep{Wicht02}. The new anelastic version has been
tested and validated against different compressible convection and dynamo
benchmarks \citep{Jones11}. To solve the system of equations (\ref{eq:anel}),
(\ref{eq:NS}) and (\ref{eq:entropy}) in spherical coordinates $(r,\theta,\phi)$,
the mass flux $\rb \vec{u}$ is decomposed into a poloidal and a toroidal
contribution

\begin{equation}
\rb\vec{u} = \vna\times\lp\vna\times W \vec{e_r}\rp +
\vna\times Z \vec{e_r},
\label{eq:decomposition}
\end{equation}
where $W$ and $Z$ are the poloidal and toroidal potentials. $W$, $Z$, $p$ and
$s$ are then expanded in spherical harmonic functions up to degree
$\ell_{\text{max}}$ in the angular variables $\theta$ and $\phi$ and in
Chebyshev polynomials up to degree $N_r$ in the radial direction. The equations
for $W$ and $Z$ are obtained by taking the horizontal part of the divergence
and the radial component of the curl of Eq.~(\ref{eq:NS}), respectively.
A more exhaustive description of the numerical method and spectral transforms
involved in the computation can be found in \citep{Glatz1} and in
\citep{Christensen07}.

\subsection{Boundary conditions}

In all the simulations presented in this study, we have assumed constant
entropy and stress-free boundary conditions for the velocity at both limits.
Under the anelastic approximation, the non-penetrating stress-free boundary
conditions is modified compared to the usual Boussinesq one as it now involves
the background density

\begin{equation}
 \dfrac{\partial}{\partial r}\lp \dfrac{1}{r^2\rb}\dfrac{\partial W}{\partial
r} \rp = \dfrac{\partial}{\partial r}\lp \dfrac{1}{r^2 \rb} Z \rp = 0
\quad\text{for}\quad r=[r_i, r_o].
\end{equation}
The non-penetrating condition of the radial motion is not affected by the
anelastic approximation and remains $u_r=0$ (or $W=0$) at both limits.

\section{Physical properties of compressible convection}

\label{sec:flowprop}

We have fixed the aspect ratio  $\eta=0.6$ to the lower value suggested for the
molecular to metallic hydrogen transition in Jupiter ($0.85~R_J$) and Saturn
($0.6~R_S$) \citep[e.g.][]{Liu08}. This limits the numerical difficulties
associated with thinner shells and therefore allows to reach higher density
stratification. The Ekman number is
kept fixed at $\text{E}=10^{-4}$, which is larger
than the most extreme values chosen by \cite{Jones09} but allows us to conduct
a large number of simulations. Following previous Boussinesq
studies \citep[e.g.][]{Christensen02}, the Prandtl number is set to $1$, while
the polytropic index is $m=2$ following \cite{Jones09}.
Being mostly interested in the effects of density stratification on zonal flow
generation, we have performed simulations at five different values of $N_\rho$
$(10^{-2}, 1, 2, 3, 4, 5)$ that span the range from nearly Boussinesq to a
density contrast of 150. In the gas giants, the density jump between the 1~bar 
level
and the bottom of the molecular region corresponds to $N_\rho\simeq
8.5$ \citep[e.g][]{Guillot99,Nettelmann08}. For numerical reasons, we cannot
afford to increase $N_\rho$ beyond $5$. However, since the density gradient
decreases rapidly with depth, $N_\rho=5$ covers already $99\%$ of the molecular
envelope when starting at the metallic transition.
The remaining rather thin upper atmosphere (roughly 500~kms) may involve
additional physical effects such as radiative transfer, weather effects or
insolation, that we do not include in our model.
For each of the stratifications used in this study, the Rayleigh number
has been varied, starting from close to onset up to 140 times the critical
value. Altogether more than 140 simulations have been computed, each running
for at least 0.2 viscous time ensuring that the nonlinear saturation has
been reached (see Table~\ref{tab:simus}).

The numerical resolution increases from $(N_r=73,\ \ell_{\text{max}} = 85)$
for Boussinesq runs close to onset to $(N_r=161,\ \ell_{\text{max}} = 341)$
for the more demanding highly supercritical simulations with strong
stratification. In the latter cases, we have resolved to imposing a two-fold or
a four-fold azimuthal symmetry, effectively solving for only half or a quarter
of the spherical shell, respectively. A comparison of testruns with or without
symmetries showed no significant statistical differences.

\subsection{Onset of convection}

\begin{table}
\centering
 \caption{Values of the critical Rayleigh numbers (and flux-based counterparts)
and azimuthal wavenumbers for different background density stratifications (with
$\text{E} = 10^{-4}$ and $\text{Pr}=1$).}
 \begin{tabular}{ccccc}
 \toprule
 $N_\rho$ & $\text{Ra}_{\text{crit}}$ & $\text{Ra}_q^*$ & $\langle\text{Ra}_q^*
\rangle_\rho$ & $m_{\text{crit}}$ \\
 \midrule
 $10^{-2}$ & $1.739\times10^{5}$ & $1.739\times10^{-7}$ & $2.666\times10^{-7}$ &
 21 \\
 1 & $5.175\times10^{5}$ & $5.175\times10^{-7}$ & $8.275\times10^{-7}$ & 34 \\
 2 & $1.141\times10^{6}$ & $1.141\times10^{-6}$ & $1.496\times10^{-6}$ & 53 \\
 3 & $1.529\times10^{6}$ & $1.529\times10^{-6}$ & $1.326\times10^{-6}$ & 72 \\
 4 & $1.852\times10^{6}$ & $1.852\times10^{-6}$ & $9.023\times10^{-7}$ & 83 \\
 5 & $2.341\times10^{6}$ & $2.341\times10^{-6}$ & $5.727\times10^{-7}$ & 93 \\
 \bottomrule
 \end{tabular}
 \label{tab:critical}
\end{table}

\begin{figure*}
 \centering
 \includegraphics[width=14cm]{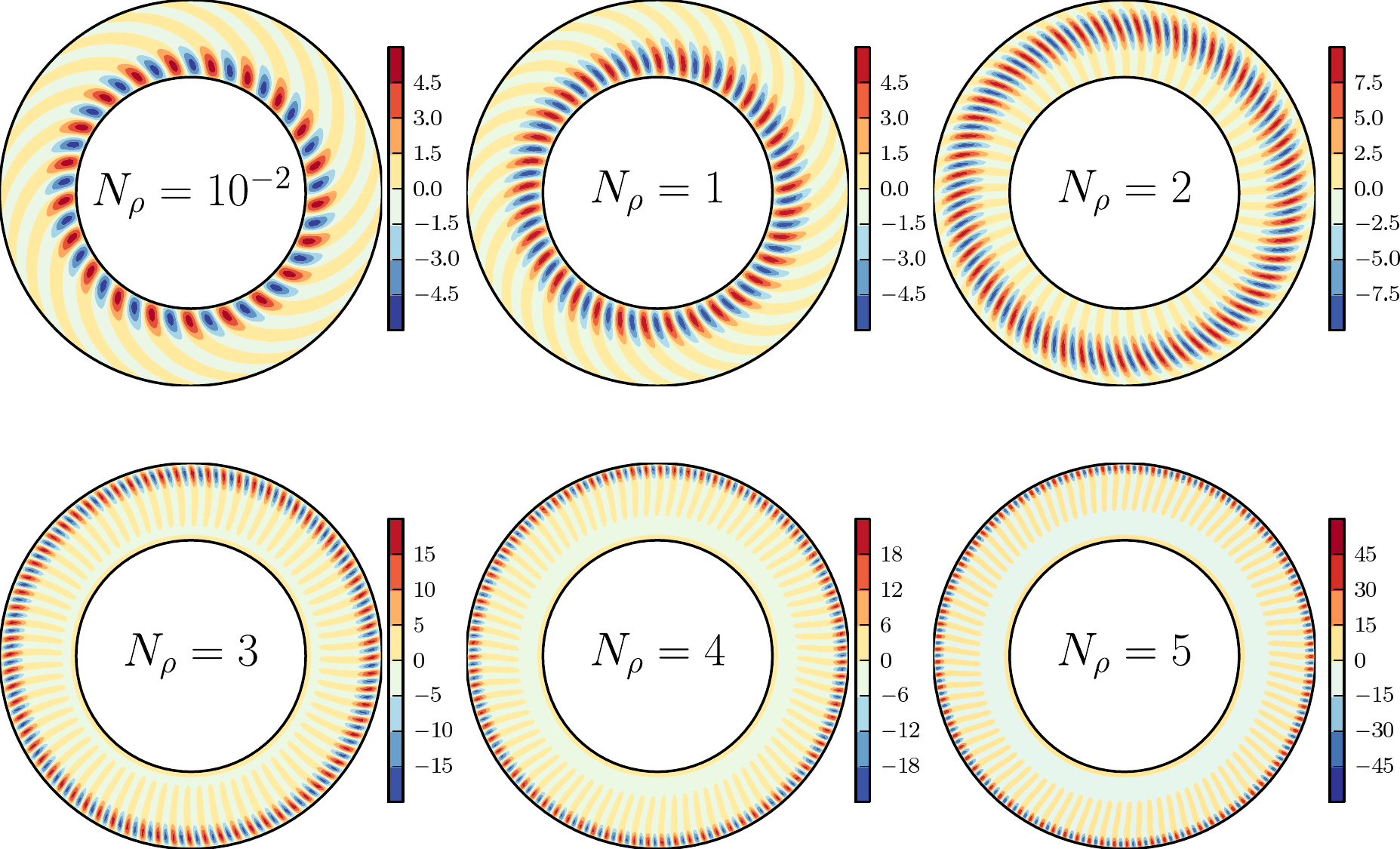}
 \caption{Radial component of the velocity $u_r$ in the equatorial plane for
different density stratifications very close to onset of convection (less than
$5\%$ over the critical Rayleigh number). Outward flows are rendered in red,
inward flows in blue, dimensionless radial velocities are expressed in
terms of Reynolds numbers.}
\label{fig:critical}
\end{figure*}

The linear stability analysis by \cite{Jones09a} shows that density
stratification  can have a significant influence both on the value of the
critical Rayleigh number and on the location of the first unstable mode.
To facilitate the comparison
between different density contrasts, we have therefore computed the
corresponding critical Rayleigh numbers (Tab.~\ref{tab:critical}). The values
have been obtained by trial and error, monitoring the growth or decay of
initial disturbances at various Rayleigh numbers.

Figure~\ref{fig:critical} emphasises how the location of the convective
instability changes when $N_\rho$ is increased. In the nearly Boussinesq case
($N_\rho=10^{-2}$), convection sets in near the tangent cylinder.
The convective columns show a strong prograde spiralling, typical for
rapidly rotating convection at a moderate Prandtl
number \citep[e.g.][]{Zhang92}. For $N_\rho=1$, the wavenumber increases but
the instability stays attached to the inner boundary. However, when increasing
$N_\rho$ further, it moves outward and is progressively confined to the very
thin outermost region.

This behaviour can be explained by the depth-dependent of buoyancy expressed
via the modified Rayleigh number (Eq.~\ref{eq:raloc}).
Figure~\ref{fig:critical} demonstrates how larger buoyancy values are more and
more confined close to the outer boundary as $N_\rho$ grows. This implies a
decrease in radial lengthscale, which goes along with a rapid increase in
wavenumber. This allows the instability to maintain a more or less spherical
cross section \citep{Glatz2, Jones09a}.

\subsection{Convection regimes}

\begin{figure*}
 \centering
 \includegraphics[width=14cm]{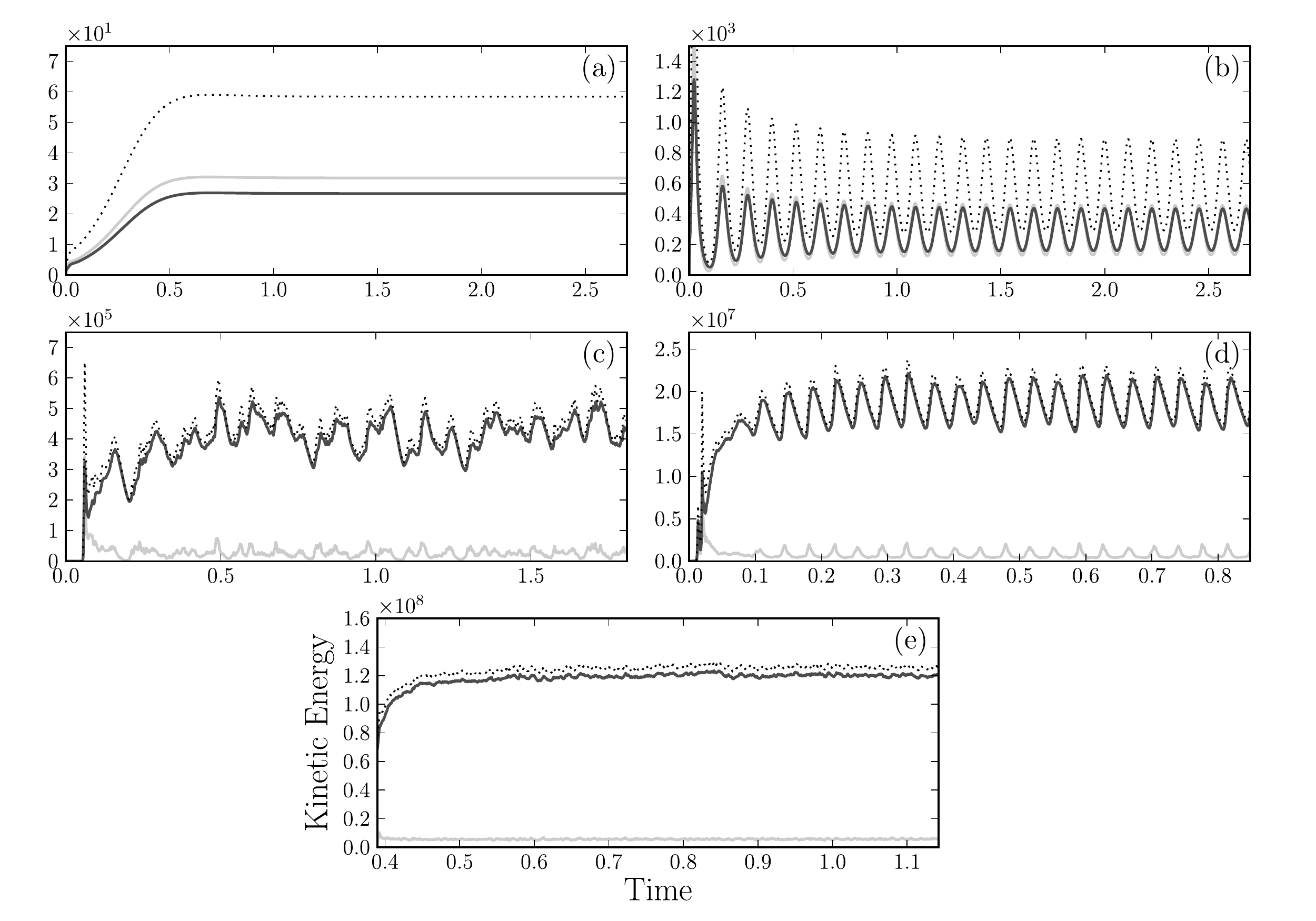}
\caption{Time series of the kinetic energy for different simulations
with a moderate density contrast ($N_\rho=2$). In each panel, the dotted black
line corresponds to the total
kinetic energy, while solid black line and solid gray line correspond
respectively to toroidal an poloidal parts of the kinetic energy.
The simulation displayed in (\textbf{a}) corresponds to
$\text{Ra}=1.145\times10^{6}$ ($1.004\times$ supercritical), (\textbf{b}) to
$\text{Ra}=1.2\times10^{6}$ ($1.05\times$ supercritical), (\textbf{c}) to
$\text{Ra}=2.2\times10^{6}$ ($1.93\times$ supercritical), (\textbf{d}) to
$\text{Ra}=8\times10^{6}$ ($7\times$ supercritical) and (\textbf{e}) to
$\text{Ra}=2\times10^{7}$ ($17.5\times$ supercritical). These five simulations
can be localized in the regime diagram displayed in
Fig.~\ref{fig:phase}.}
\label{fig:ts}
\end{figure*}

The properties of rotating thermal convection have been explored extensively
under the Boussinesq approximation \citep[e.g.][]{Zhang92,Grote01,Christensen01}.
The respective studies show that when the Rayleigh number is increased beyond onset
distinct regimes with different temporal behavior are encountered.
Here we explore whether these regimes are retained in the presence
of density stratification.

\begin{figure}
 \centering
   \includegraphics[width=8.8cm]{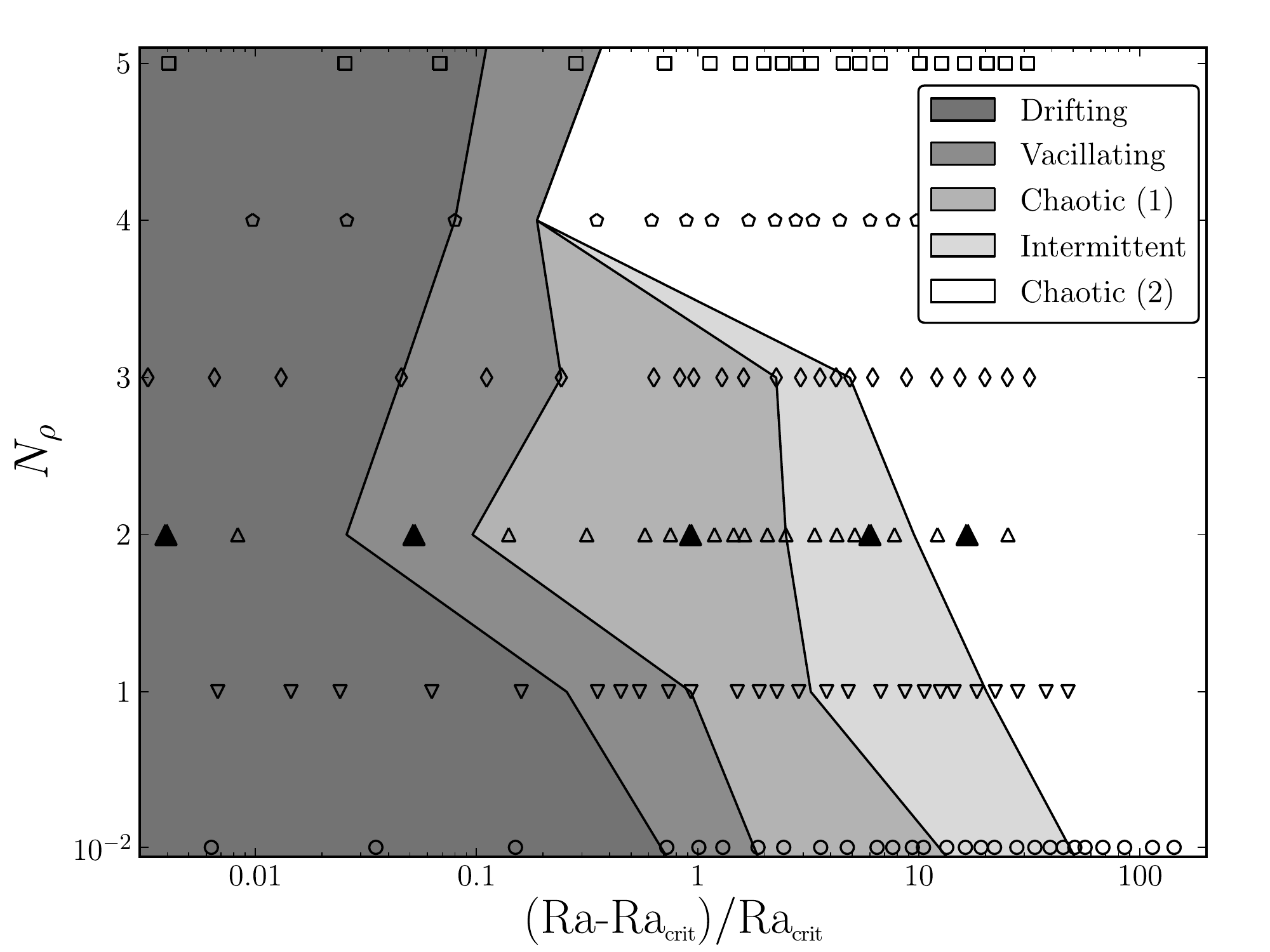}
\caption{Regime diagram $(\text{Ra}, N_\rho)$ displaying how the different
nonlinear regimes of rotating compressible convection depend on supercriticality
and density stratification. Each symbol corresponds to one
simulation computed in this study. The five black triangles
denote the localisations of the five simulations emphasised in
Fig.~\ref{fig:ts}.}
\label{fig:phase}
\end{figure}

Figure~\ref{fig:ts} illustrates the time dependence in the temporal variation
characteristic for the different regimes at $N_\rho=2$ and
in Fig.~\ref{fig:phase} we attempt to show how the regimes boundaries change
on increasing $N_\rho$. At $N_\rho=2$, we still find the same regimes identified
in the Boussinesq simulations. Very close to onset, the solutions simply drift
in azimuth due to its Rossby wave character \citep{Busse94} and the kinetic
energy becomes stationary (Fig.~\ref{fig:ts}a). The drift is a consequence of
both the height change of the container and the background density
stratification encountered by the convective instability. Its amplitude and
direction depends on the chosen stratification \citep{Evonuk08, Glatz09}. At a
still rather small supercriticality, the solution starts to vacillate in
amplitude, while retaining its columnar form. Figure~\ref{fig:ts}b demonstrates
that the poloidal and toroidal kinetic energies oscillate in phase and have a
comparable magnitude.

A further increase of the Rayleigh number leads to a third regime with a more
chaotic time-dependence (Fig.~\ref{fig:ts}c). The zonal flow production is
already so efficient that the toroidal energy dominates the kinetic energy
budget. Since the poloidal flow is directly driven by convective up- and
downwellings, its amplitude can serve as a proxy for the convective flow vigor.
Zonal flows are the axisymmetric contribution of toroidal flow and already
dominate the toroidal energy here. Both the convective and the zonal flows
display large amplitude modulations. A fourth regime with intermittent nearly
oscillatory convection is encountered once $\text{Ra} \gtrsim
4\times\text{Ra}_c$ (panel (d) in Fig.~\ref{fig:ts}).
Here, the zonal flow periodically becomes so large that the associated
shear disrupts the convective structures \citep[see for
further details][]{Christensen01,Grote01,Simitev03,Heimpel12}. The Reynolds
stresses then decrease and the zonal flow cannot be maintained any more against
viscous forces. The convection recovers once the zonal flows amplitude has
decreased sufficiently. A similar but less regular competition may also cause
the large variation in the chaotic regime. Typical for this intermittent regime,
which has mainly been observed in Boussinesq simulations but has also been
reported in some anelastic simulations of turbulent convection in solar-like
stars \citep{Ballot07}, is that the convective features typically occupy only an
azimuthal fraction of the volume. Finally, in the strongly supercritical regime,
convection becomes once more chaotic but the variations are much smaller than in
the first chaotic regime (Fig.~\ref{fig:ts}e). Convective features now fill the
whole spherical shell and efficiently drive strong zonal winds.

For $N_\rho=2$ we thus find the following regime succession also typical for 
Boussinesq convection at $\text{Pr} \lesssim 1$ \citep{Grote01}:

\[
 \text{Drifting}  \rightarrow \text{Vacillating} \rightarrow
\text{Chaotic} \rightarrow \text{Intermittent} \rightarrow
\text{Chaotic}\;.
\]
The regime diagram displayed in Fig.~\ref{fig:phase} shows that the
first chaotic
and the intermittent regime tend to vanish for larger values of $N_\rho$. For
$N_\rho=[4-5]$, we found neither the first chaotic nor the intermittent regime,
which could mean that both have retreated to a rather small region of the
parameter space or are missing altogether. Due to the lack of an intermittent
regime, however, the separation of the first and second chaotic regime is less
clear cut, being based only on the variation amplitude.  The narrowing of the
regime diagram for increasing density stratification was also found in
3-D simulations of convection in rapidly rotating isothermal spherical shells
\citep{Tortorella}.

\begin{figure}
 \centering
   \includegraphics[width=8.8cm]{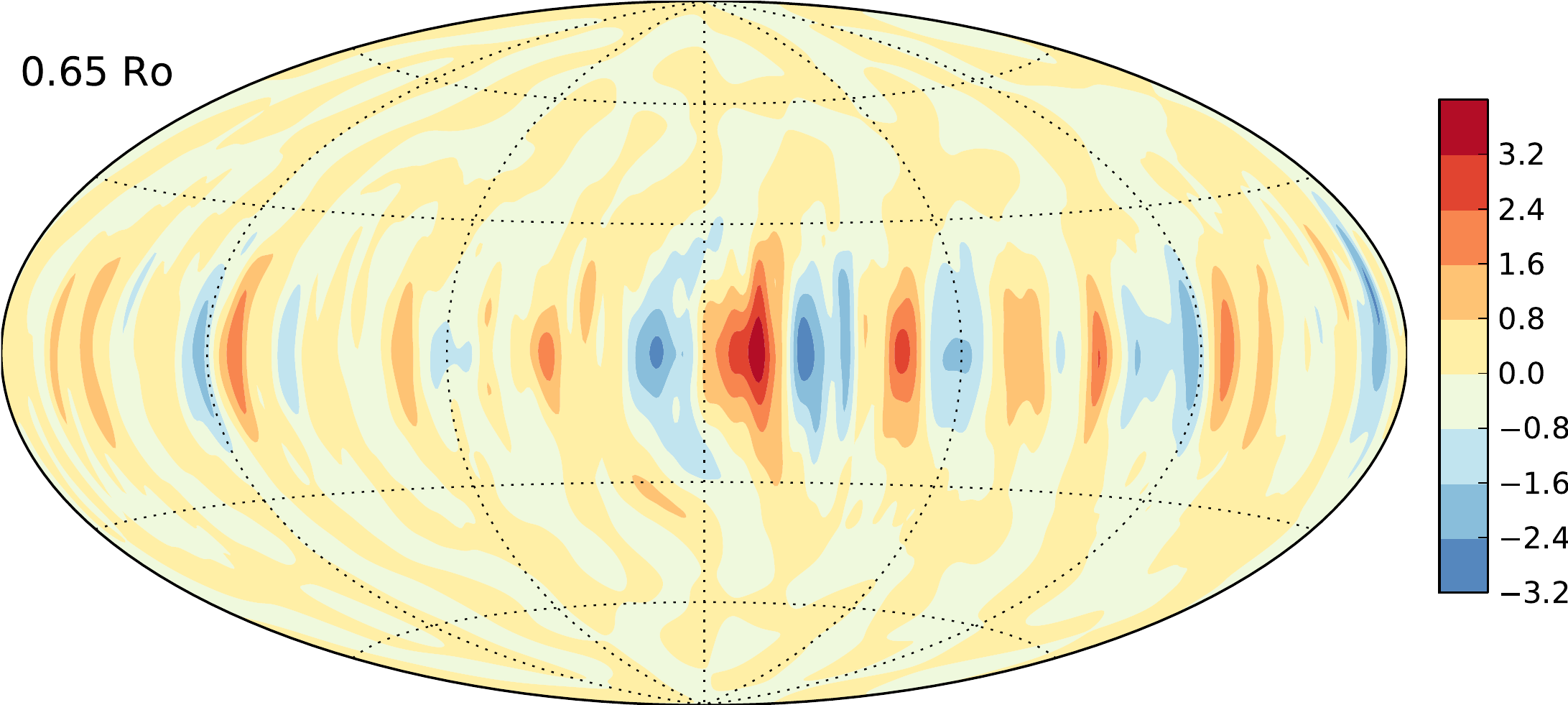}
   \includegraphics[width=8.8cm]{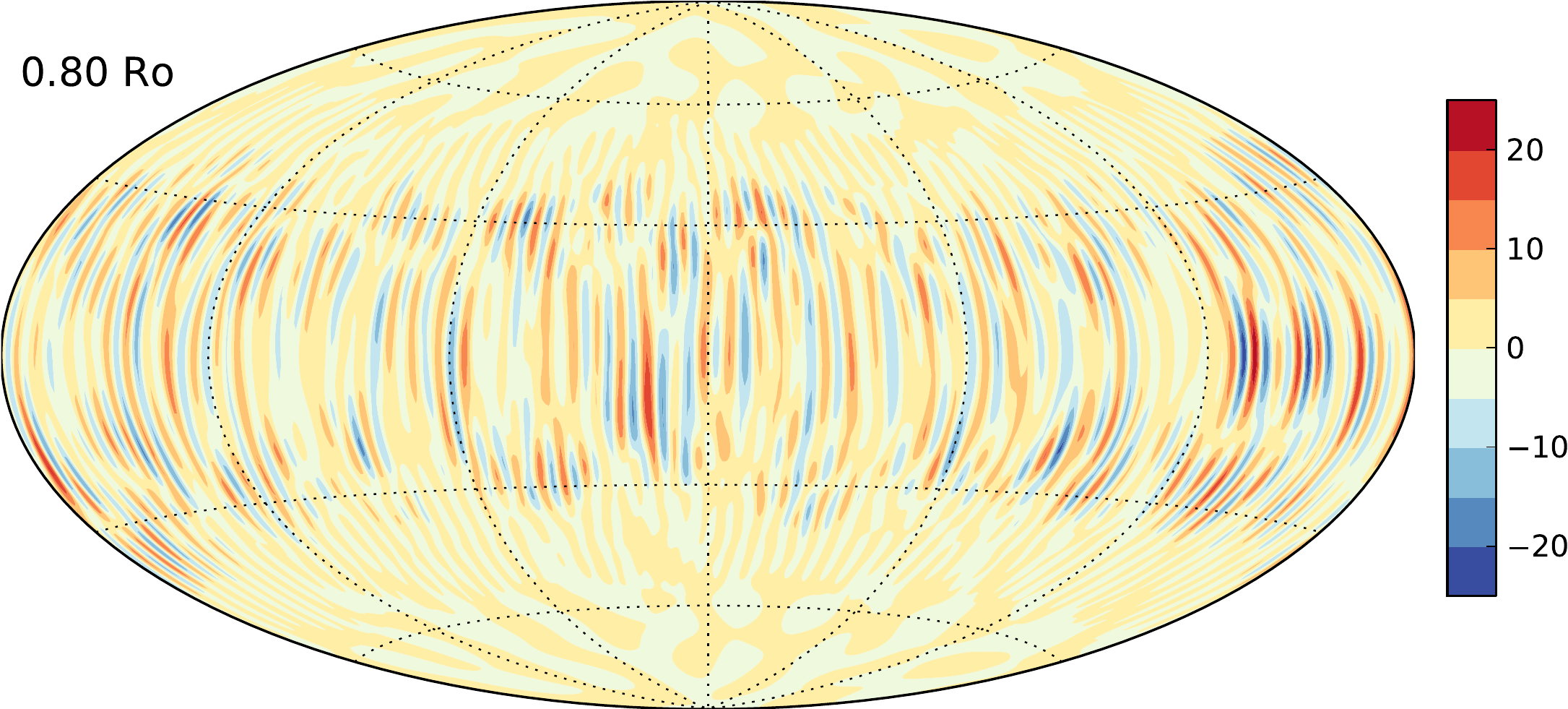}
   \includegraphics[width=8.8cm]{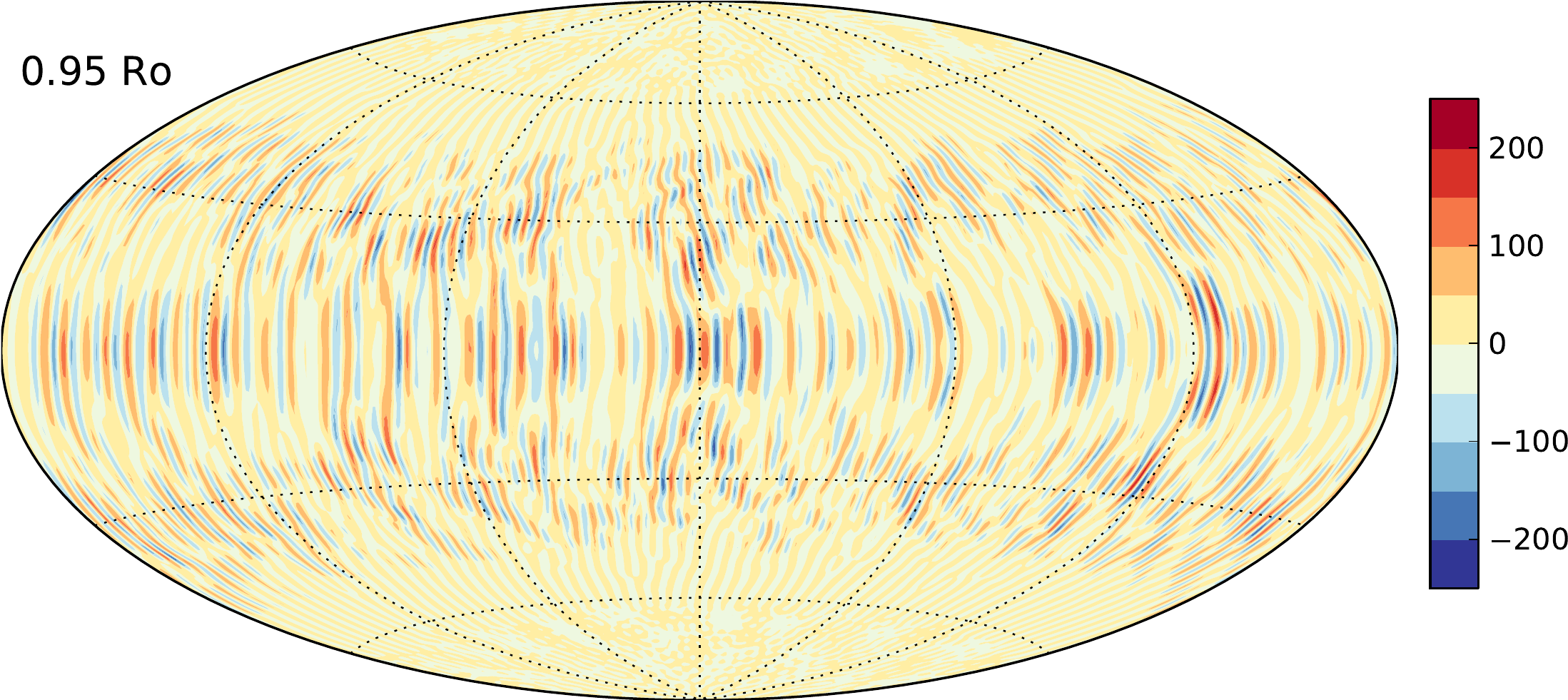}
\caption{Radial velocity $u_r$ on spherical surfaces at three different radii
close to the inner boundary ($r=0.65r_o$), at the layer middle, and close to the
outer boundary ($r=0.95r_o$). Outward flows are rendered in red, inward flows
in blue, dimensionless radial velocities are expressed in terms of
Reynolds numbers. From a model with $N_\rho=5$ and $\text{Ra}=8\times 10^{6}$.}
\label{fig:vranel}
\end{figure}

The large amplitude oscillations in the first chaotic and the intermittent
regime require a more global interaction between convection and zonal flow.
When the density stratification is increased, however, the solution becomes
more small scale and chaotic and the convective columns progressively loose their
integrity in the direction of the rotation axis \citep{Glatz09,Jones09}. This
likely prevents the required more global interaction and the related regimes at
larger density stratification.

\subsection{Flow properties}
The background density stratification causes significant changes in the 
convective flow that we discuss first before describing the zonal flows in 
section~\ref{zonal}.

\begin{figure*}
 \centering
 \includegraphics[width=15cm]{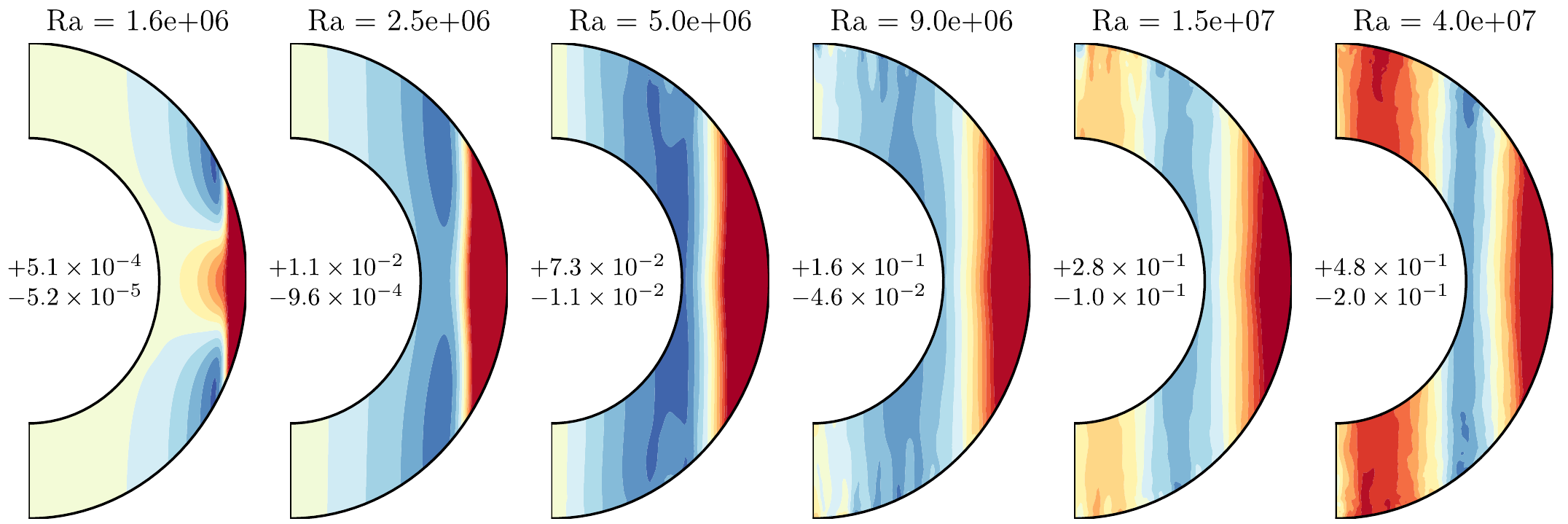}
 \caption{Zonally averaged azimuthal velocity in the meridian plane for
simulations with $N_\rho=3$ and different Rayleigh numbers. Colorscales
are centered around zero: prograde jets are rendered in red, retrograde jets in
blue. Prograde contours have been truncated in amplitude to emphasize the
structure of the retrograde flows. Dimensionless azimuthal velocities are
expressed here in terms of Rossby numbers. The extrema of the zonal flow
velocity are indicated in the center of each panel.}
\label{fig:vp_N3}
\end{figure*}

\begin{figure*}
 \centering
\includegraphics[width=15cm]{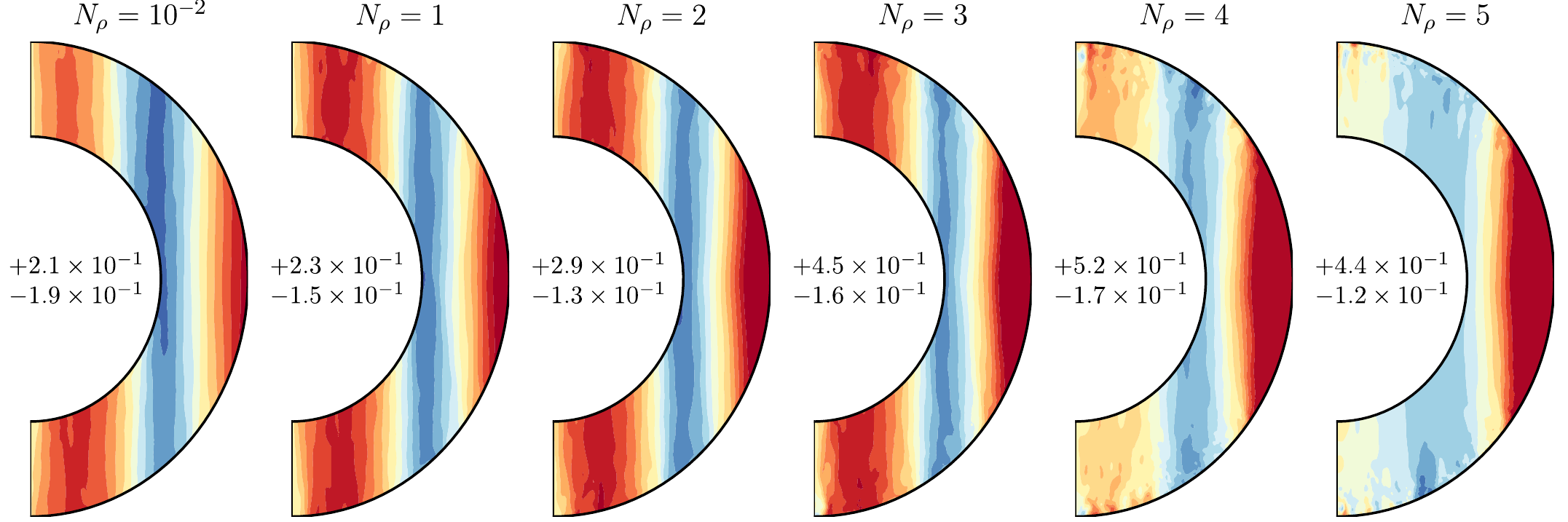}
 \caption{Zonally averaged azimuthal velocity in the meridian plane for
simulations with different density stratifications. The different setups
correspond to $N_\rho=10^{-2}$, $\langle\text{Ra}_q^*\rangle_\rho=1.70\times
10^{-4}$ (i.e. $\text{Ra}=1.2\times 10^{7}$); $N_\rho=1$,
$\langle\text{Ra}_q^*\rangle_\rho=1.50\times 10^{-4}$
(i.e. $\text{Ra}=1.2\times 10^{7}$); $N_\rho=2$, 
$\langle\text{Ra}_q^*\rangle_\rho=1.37\times 10^{-4}$
(i.e. $\text{Ra}=1.5\times 10^{7}$); $N_\rho=3$, 
$\langle\text{Ra}_q^*\rangle_\rho=2.63\times 10^{-4}$
(i.e. $\text{Ra}=3.2\times 10^{7}$);$N_\rho=4$, 
$\langle\text{Ra}_q^*\rangle_\rho=2.84\times 10^{-4}$
(i.e. $\text{Ra}=6\times 10^{7}$) and $N_\rho=5$, 
$\langle\text{Ra}_q^*\rangle_\rho=1.48\times 10^{-4}$
(i.e. $\text{Ra}=7.5\times 10^{7}$) .
Colorscales are centered around zero: prograde jets are rendered
in red, retrograde jets in blue. Prograde contours have been truncated in
amplitude to emphasize the structure of the retrograde flows. Dimensionless
azimuthal velocities are expressed here in terms of Rossby numbers. The extrema 
of the zonal flow velocity are indicated in the center of each panel.}
\label{fig:vp_rho}
\end{figure*}

\subsubsection{Convective flows}

Figure~\ref{fig:vranel} illustrates how the radial velocity varies with depth 
in a strongly-stratified anelastic simulation at $N_\rho=5$. The Rayleigh 
number of $\text{Ra}=8\times10^6$ guarantees that the solution is safely located
in the second chaotic regime. Deeper within the interior, the convection is
characterized by larger azimuthal length scales and
lower amplitudes. With increasing radius, the azimuthal length scale decreases,
while the amplitude increases. The convective features remain roughly aligned
with the rotation axis at any radius. The changes in length scale are coupled to
the  background density profile as the density
scale height defined by $H_\rho = -(d\ln\rb/dr)^{-1}$ is a good proxy for the
typical size of the convective eddies. At $N_\rho=5$, this scale height
decreases by a factor of five when going from the inner to the outer boundary,
which about explains the variation shown in Fig.~\ref{fig:vranel}.

The increase in amplitude can be explained by the variation of the
background density: since $\rb$ decreases by a factor
of about $150$ from $r_i$ to $r_o$, a rising (falling) convective
plume compensates this with increasing (decreasing) flow amplitude.
Figure~\ref{fig:vranel} shows an increase of about 60 which seems somewhat on 
the low side. However, the radial increase in the number of convective columns
and the scale change show that the convective features hardly reach through the
whole shell without being modified.

\subsubsection{Zonal flows}
\label{zonal}

\cite{Evonuk08} and \cite{Glatz09} argue that the density stratification 
can significantly influence the generation of zonal winds. Exploring the 
possible related effects is the main focus of our study.
Figure~\ref{fig:vp_N3}  shows how the zonal flows evolve on increasing the
Rayleigh number in moderately stratified simulations ($N_\rho=3$).
Close to the onset of convection, a thin prograde equatorial jet develops in the
very outer part of the shell. The latitudinal extent of this jet is 
smaller than in Boussinesq convection because the convective columns are
confined closer to the outer boundary (see the simulation with $N_\rho=3$ in
Fig.~\ref{fig:critical}). It is flanked by weak amplitude retrograde
jets that strongly vary along $z$.
As $\text{Ra}$ is increased, convection starts to fill the whole shell, the
equatorial  jet broadens and the zonal winds become more and more
geostrophic with a second retrograde jet that reaches throughout the shell.
For $\text{Ra}=5\times 10^{6}$ ($3.3\times$ critical) the equatorial jet
has reached a maximum width that it retains at even higher Rayleigh numbers.
The retrograde jet broadens toward higher latitudes until additional flanking
prograde jets develop below and above the inner boundary at large Rayleigh
numbers and confine it to mid latitudes. The amplitude of these high-latitude
jets becomes comparable to that of the equatorial 
jet for strongly nonlinear simulations ($\text{Ra}=4\times 10^{7}$,
$26\times\text{Ra}_c$), while the retrograde jet vigor intensifies to about
50\% of this amplitude. In this multiple jets regime, retrograde zonal
winds are anchored to the tangent cylinder in a similar way to what was already
observed in Boussinesq simulations \citep[e.g.][]{Christensen01,Heimpel05}.

The strong $z$-dependence at low Rayleigh numbers is a typical thermal wind
feature, which can be understood by considering
the $\phi$-component of the
curl of the momentum equation (\ref{eq:NS}):
\begin{equation}
\begin{aligned}
 \dfrac{D\omega_\phi}{Dt} &=& r\sin\theta\ (\vec{\omega}\cdot\vec{\nabla}
)\lp\dfrac{u_\phi}{r\sin\theta}\rp -\omega_\phi\vec{\nabla} \cdot\vec{u} 
+\dfrac{2}{\text{E}}\dfrac {\partial u_\phi}{\partial z} \\
& &-\dfrac{\text{Ra}}{\text{Pr}}\dfrac{r_o^2}{r^3}\dfrac{\partial
s}{\partial\theta}
+\left[\vec{\nabla} \times
\lp\rb^{-1}\vec{\nabla}\cdot\tens{S}\rp\right]_\phi.
\end{aligned}
\end{equation}
Here $D/Dt$ corresponds to the substantial time derivative and $\omega_\phi$ is
the azimuthal vorticity component.
For low Rayleigh numbers, inertial contributions can be neglected, and the
$z$-dependence of the zonal flows is ruled by the thermal wind balance:
\begin{equation}
\dfrac{2}{\text{E}}\dfrac{\partial \overline{u_\phi}}{\partial z} \simeq
\dfrac{\text{Ra}}{\text{Pr}}\dfrac{r_o^2}{r^3}\dfrac{\partial
\overline{s}}{\partial\theta},
\label{eq:thermalwind}
\end{equation}
where overlines denote an azimuthal average.
This allows us to define the typical lengthscale 
\begin{equation}
 L_z = \lp \dfrac{\partial \ln \overline{u_\phi}}{\partial z} \rp^{-1}
 =\overline{u_\phi}\ \lp \dfrac{\text{Ra}\ \text{E}}{2
\text{Pr}}\dfrac{r_o^2}{r^3}\dfrac{\partial
\overline{s}}{\partial\theta} \rp^{-1}.
\end{equation}
For vanishing Ekman numbers, $L_z$ approaches infinity and zonal flows become
independent of $z$, recovering Taylor-Proudman theorem and therefore a
geostrophic structure. For the lowest Rayleigh number solution shown in
Fig.~\ref{fig:vp_N3}a, the thermal winds caused $z$-variation is still clearly
apparent where $u_\phi$ is weak and $L_z$ therefore small. At larger Rayleigh
numbers, the thermal wind balance still holds in a statistical sense. According
to Eq.~(\ref{eq:thermalwind}), $\partial \overline{u_\phi}/\partial z$ increases
linearly with Ra, while, at least closer to onset, the zonal flow
amplitude rises faster. $L_z$ therefore grows
with the Rayleigh number and the $z$-variation is gradually lost.
In the simulations of \cite{Kaspi09} however, the zonal winds still vary
strongly close to the surface. These are likely promoted by the depth-dependence
of the thermal expansion coefficient, which is neglected in our ideal gas 
model.

Figure~\ref{fig:vp_rho} compares the zonal flows at different density 
stratifications for simulations with similar $\langle
\text{Ra}_q^*\rangle_\rho$ and demonstrates that the structure changes very
little for $N_\rho\le 3$. In the two cases at $N_\rho=4$ and $N_\rho=5$ the
high-latitude jets are still little pronounced and the retrograde jet is also
rather weak. According to the development shown in Fig.~\ref{fig:vp_N3} they
seem to correspond to a lower supercriticality.  The fact that $N_\rho=4$ and
$5$ simulations have a supercriticality that is about three times higher than
that of the $N_\rho=2$ case, however, implies that significantly
larger supercriticalities are required to reach a comparable multiple jet
structure when the density stratification is strong. Since both large Rayleigh
numbers and strong density stratifications promote smaller scales, we could not
afford to further increase $\text{Ra}$ for $N_\rho>3$. While the zonal winds are
very geostrophic for smaller density stratification, additional non-geostrophic
small scale features start to appear for larger stratifications as it has
already been reported by \cite{Jones09}. They are strongly time-dependent and
therefore disappear when time averaged quantities are considered.

Except for these secondary features and the requirement for larger supercriticalities 
at stronger stratifications, the zonal flow structure is surprisingly
independent of the background density profile. We now turn to discussing their
amplitude.

\section{Scaling laws}
\label{sec:scaling}

To evaluate how the different flow properties depend on the 
background density stratification, we consider time averaged quantities in the
following.
Each simulation has been run for at least $0.2$ viscous diffusions times to get rid of any transients
and to allow for an averaging period long enough to suppress short term
variations. The Reynolds numbers Re and Rossby numbers $\text{Ro}=\text{E}\
\text{Re}$, used to quantify the amplitude 
of different flow contributions are based on the non-dimensional
time averaged root-mean square velocity: 

\begin{equation}
  \text{Re} =\left(\langle u^2\rangle\right)^{1/2},
\end{equation}
with

\begin{equation}
\langle u^2 \rangle =\dfrac{1}{\tau}\dfrac{1}{V_s} \int_{t_0}^{t_0+\tau}\int_V
\vec{u}^2 dt\ dV,
\end{equation}
where $\tau$ is the averaging interval, $V_s$ is the volume of the spherical
shell and $dV$ is a volume element.
As pointed out by \cite{Kaspi09}, a comparison of simulations with
different density backgrounds may be more relevant if one considers
mass-weighted average quantities. Therefore, the mass-weighted counterparts of
Reynolds and Rossby numbers are defined using the time averaged kinetic
energy, that is

\begin{equation}
  \langle\text{Re}\rangle_\rho
=\left(\frac{2E_{\text{kin}}}{\langle\rb\rangle}\right)^{1/2},
\end{equation}
with

\begin{equation}
E_{\text{kin}} =\dfrac{1}{\tau}\dfrac{1}{V_s} \int_{t_0}^{t_0+\tau}\int_V
\rb\vec{u}^2 dt\ dV.
\end{equation}

\begin{figure*}
 \centering
 \includegraphics[width=12cm]{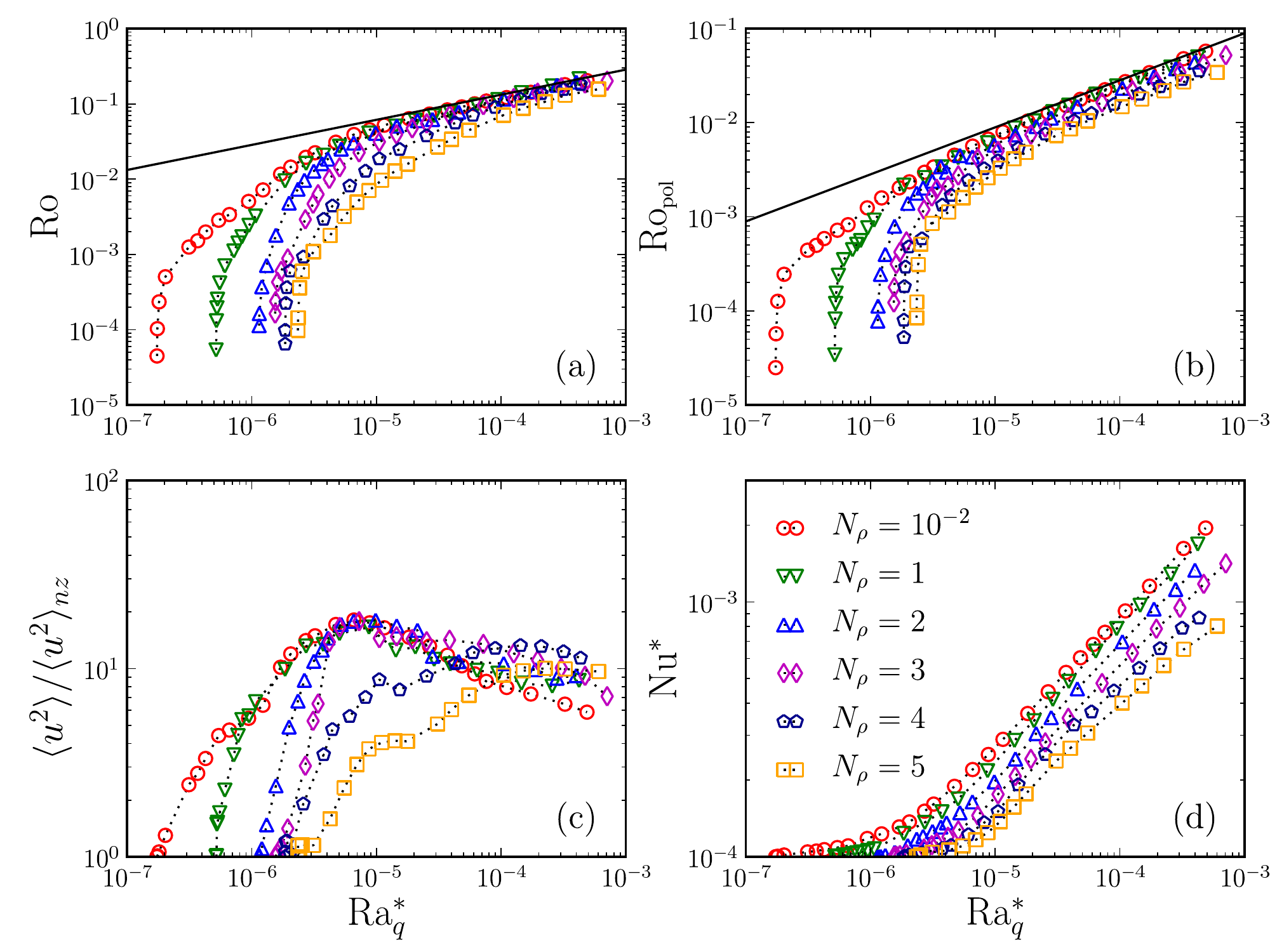}
 \caption{Time-averaged properties plotted against flux-based Rayleigh number
defined in Eq.~(\ref{eq:fluxRa}): (a) Rossby number; (b) Rossby number for the
poloidal flow component; (c) ratio of square velocity over non-zonal square
velocity; (d) modified Nusselt number.}
\label{fig:ranu}
\end{figure*}

\begin{figure*}
 \centering
 \includegraphics[width=12cm]{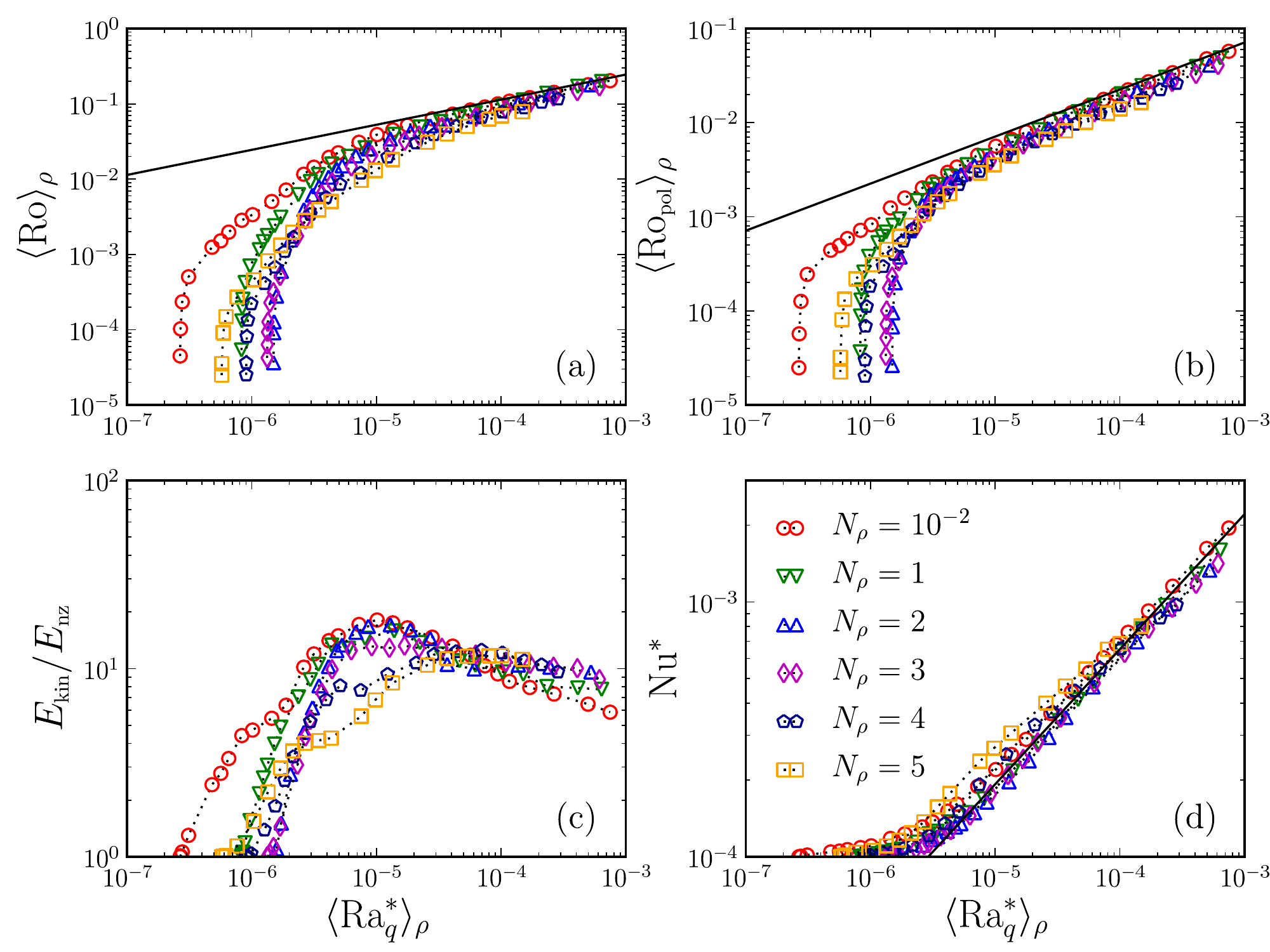}
 \caption{Time-averaged properties plotted against mass-weighted flux-based
Rayleigh number defined in Eq.~(\ref{eq:fluxRarho}): (a) mass-weighted Rossby
number; (b) mass-weighted Rossby number for the poloidal flow component; (c)
ratio of kinetic energy over non-zonal kinetic energy; (d) modified Nusselt
number.}
\label{fig:resc_ranu}
\end{figure*}

Figures~\ref{fig:ranu} and \ref{fig:resc_ranu} show how different time-average
properties for the various density stratifications explored here 
depend on the flux-based Rayleigh number (Eq.~\ref{eq:fluxRa})
and the mass-weighted flux-based Rayleigh number (Eq.~\ref{eq:fluxRarho}), respectively.
The total Rossby number Ro (Fig.~\ref{fig:ranu}a) and the poloidal Rossby
number $\text{Ro}_{\text{pol}}$ (Fig.~\ref{fig:ranu}b), as well as their
mass-weighted counterparts (Fig.~\ref{fig:resc_ranu}a-b) first rise steeply. The
slope flattens once convection fills the whole shell and seems to reach an
asymptotic value for large Rayleigh numbers.  The asymptotic slopes are larger
for the poloidal than for the total flow amplitude. In both
Figs.~\ref{fig:ranu} and \ref{fig:resc_ranu}, the curves collapse at larger
Rayleigh numbers, and the asymptotic slopes become independent of the density
stratification. This is even more convincing when mass-weighted average
quantities are considered, which basically confirms \cite{Kaspi09}.
The remaining differences at low Rayleigh numbers can at least partly be
explained by the different critical values. Compared to the gas giants, the 
Rossby numbers where our simulations reach
the asymptotic regime are roughly one order of magnitude larger. However this
is likely  a consequence of the moderate Ekman number used in our models as
\cite{Christensen02} showed that this transitional Rossby number decreases with
Ekman number.

Figure~\ref{fig:resc_ranu}c displays the ratio of the total to the non-zonal
kinetic energy, while Fig.~\ref{fig:ranu}c corresponds to the ratio of square
velocities only. For all density stratifications, this ratio initially increases
in the weakly supercritical regime but decays once the convection becomes
strongly nonlinear. Similarly to what has been reported for Boussinesq
simulations \citep{Christensen02}, the maximum is reached close to the
transition to the intermittent regime (see Fig.~\ref{fig:phase} for
$N_\rho\le3$). 

In the nearly Boussinesq simulations, the maximum ratio is 
$E_{\text{kin}}/E_{\text{nz}}\simeq 20$, which means that $95$\% of the energy
are carried by zonal flows.  For $N_\rho=5$ this decreases to about $90$\% at a 
maximum ratio of $E_{\text{kin}}/E_{\text{nz}}\simeq 10$.
These values are on the low side of the expected ratio
for Jupiter: around $50-100$ from the work of \cite{Salyk06}. However, this
moderate maximum ratio is  once more a consequence of the too large Ekman number
used in our simulations \citep{Christensen02}.
The slope of the decrease for larger Rayleigh numbers also seems to vary 
with $N_\rho$, but a clear dependence is difficult to extract from our dataset.
Considering mass-weighted quantities slightly helps to merge the different
simulations, but they are far from collapsing as nicely as for the Rossby
numbers. 

Panels (d) in Figs.~\ref{fig:ranu} and \ref{fig:resc_ranu} show the dependence 
of the modified Nusselt number on the Rayleigh numbers. Close to onset of 
convection, the heat transport is dominated by conduction. The classically 
defined Nusselt number remains close to unity and the Nusselt number increases 
only weakly with the Rayleigh number. When the convective motions become more 
vigorous, convective heat transport starts to dominate and the Nusselt number 
rises steeply.  Using the mass-weighted Rayleigh
number $\langle\text{Ra}_q^*\rangle_\rho$ allows to collapse the curves for
different  density stratifications and a similar asymptotic slope is reached. 
This once more emphasises that mass-weighted quantities are probably more
meaningful parameters, when considering strongly nonlinear convection. 

Based on the results displayed in Fig.~\ref{fig:resc_ranu} we have 
estimated the asymptotic dependence of $\langle\text{Ro}\rangle_\rho$,
$\langle\text{Ro}_{\text{pol}}\rangle_\rho$ and
$\text{Nu}^*$ on $\langle\text{Ra}_q^*\rangle_\rho$ to:
\begin{equation}
 \langle\text{Ro}\rangle_\rho =
2.45\left(\langle\hbox{Ra}_q^*\rangle_\rho\right)^{1/3},
 \label{eq:rossby_scale}
\end{equation}

\begin{equation}
 \langle\text{Ro}_{\text{pol}}\rangle_\rho =
2.25\left(\langle\hbox{Ra}_q^*\rangle_\rho\right)^{1/2},
\label{eq:roconv_scale}
\end{equation}

\begin{equation}
 \text{Nu}^* = 0.086\left(\langle\hbox{Ra}_q^*\rangle_\rho\right)^{0.53}.
\label{eq:nu_scale}
\end{equation}
Black lines in Fig.~\ref{fig:resc_ranu} show the respective asymptotes. The same
slopes have been obtained for the usual Rossby and poloidal Rossby numbers in
Fig.~\ref{fig:ranu}, as the nearly Boussinesq simulations hardly depend on the
density background (therefore mass-weighted counterparts of Rossby and
poloidal Rossby numbers are equivalent to their usual definitions).

The amplitude of the convective flow, quantified via the poloidal Rossby number, 
is thus found to depend on the mass-weighted flux-based Rayleigh
number with a slope of $1/2$. This is compatible with both laboratory
experiments of turbulent rotating convection \citep[e.g.][]{Fernando91} and
previous anelastic numerical simulations \citep{Showman10}
but slightly differs from the exponent $2/5$ suggested by \cite{Christensen02}. 
Investigating the reason for this difference would probably require more simulations 
at lower Ekman numbers and lies beyond the scope of this study.

The Nusselt number scaling Eq.~(\ref{eq:nu_scale}) is in a very good agreement
with previous laws derived by \cite{Christensen02} and \cite{Christensen06} for 
Boussinesq simulations. In the highly supercritical regime, the convective
transport dominates diffusion. Then $\text{Nu} \sim u_{\text{conv}}\ \delta s$,
or $\text{Nu}^*\sim \text{Ro}_{\text{pol}}\ \delta s$ where $\delta s$ is a
measure for the typical entropy fluctuations. Since $\text{Nu}^*$ and
$\text{Ro}_{\text{pol}}$ depend with a very similar exponent on
$\langle\text{Ra}_q^*\rangle_\rho$, the typical entropy
fluctuations $\delta s$ may vary very little with $\langle
\text{Ra}_q^*\rangle_\rho$ in the asymptotic regime, in agreement with
previous Boussinesq results \citep{Christensen02}.

In the highly supercritical regime, the majority of the kinetic energy is
carried by zonal flows. The asymptotic law for the total Rossby number
(Eq.~\ref{eq:rossby_scale}) therefore applies to the amplitude of zonal winds.
Based on energy considerations for the highly supercritical regime,
\cite{Showman10} derived a scaling law for strong zonal jets that suggest an
exponent of $1/2$ rather than the $1/3$ promoted here. We further discuss this
difference in the following section.

\subsection{Force balance and correlation of the convective flow}

\begin{figure}
 \centering
  \includegraphics[width=8cm]{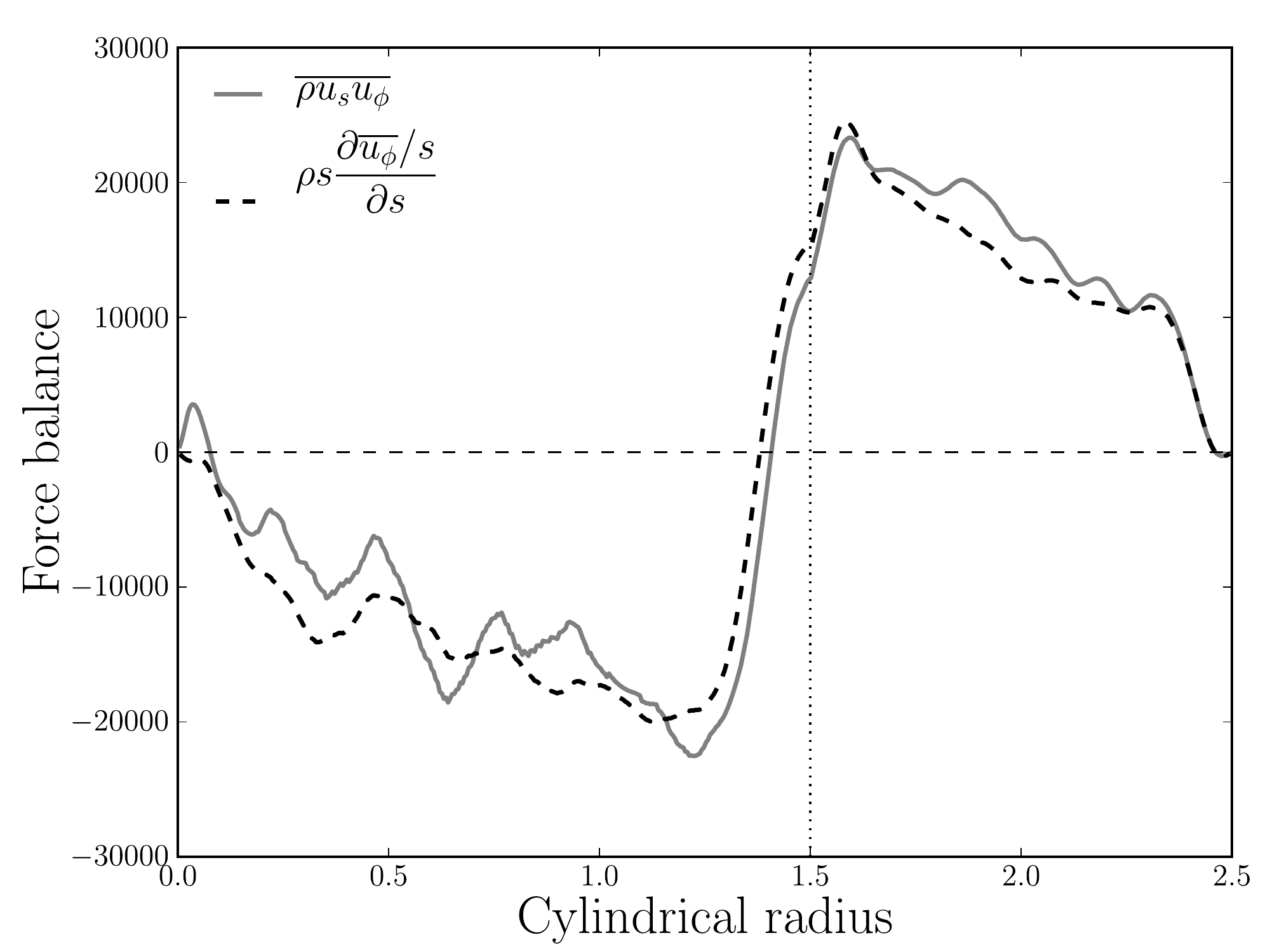}
 \caption{Time average of Reynolds stresses (solid grey line) and viscous force
(dashed black line) integrated over cylinders (Eq.~\ref{eq:balance}) for a
model with $N_\rho=3$ and $\text{Ra}=2\times10^7$. The vertical line corresponds
to the tangent cylinder.}
\label{fig:balance}
\end{figure}

To explore how the strong zonal flows are generated and maintained,
we analyze the azimuthal component of the Navier-Stokes equation~(\ref{eq:NS}) 
integrated over cylindrical surfaces of radius $s$. 
The Coriolis term is proportional to the net mass flux across these surfaces and 
therefore vanishes. The integrated Navier-Stokes equation thus simplifies to 
\begin{equation}
\begin{aligned}
 \displaystyle\int_S \rb \dfrac{\partial u_\phi}{\partial t} s d\phi dz =
 &-\int_S \dfrac{1}{s^2}\dfrac{\partial}{\partial s}\left[ s^2  \rb u_s
u_\phi\right] s d\phi dz \\
 &+ \int_S \dfrac{1}{s^2}\dfrac{\partial}{\partial
s}\left[ s^2 \rb s \dfrac{\partial}{\partial s} \lp \dfrac{u_\phi}{s} \rp
\right] s d\phi dz.
\end{aligned}
\end{equation}
On time average the two terms on the right hand side, Reynolds stress and viscous stress, 
should balance. This leads to the following condition 

\begin{equation}
 \displaystyle \int_z \rb s\dfrac{\partial}{\partial
s}\left(\dfrac{\overline{{u}_\phi}}{s}\right) dz = \int_z \overline{\rb  u_s
u_\phi} dz,
\label{eq:balance}
\end{equation}
where the overlines indicate an azimuthal
average. Figure~\ref{fig:balance} compares the left and right hand side of 
Eq.~(\ref{eq:balance}) for a case with $N_\rho=3$ and  $\text{Ra}=2\times 10^7$ 
(corresponding to $\langle Ra_q^*\rangle_\rho =1.7\times 10^{-4}$) that lies
very close to the asymptotic regimes found in Fig.~\ref{fig:resc_ranu}.
These two terms balance to a good degree, which proves that zonal flows are
indeed maintained by Reynolds stresses against viscous dissipation.

According to Eq.~(\ref{fig:balance}), we can also directly relate the viscous 
contribution in Fig.~\ref{fig:resc_ranu} to  zonal flow gradients. Zonal flows 
are thus expected to increase outside the tangent cylinder, located at $s=1.5$, 
and to mainly decrease inside the tangent cylinder.  A conversion to Reynolds 
stresses suggests that the strongest stresses can be found around the tangent 
cylinder and close to the outer boundary. The stresses are negative around the 
tangent cylinder and predominantly positive elsewhere.  All this corresponds 
nicely to the multiple jet cases described above with a prograde equatorial 
jet, a retrograde jet around the tangent cylinder, and a prograde, high
latitude jet in each hemisphere.

\begin{figure}
 \centering
  \includegraphics[width=8cm]{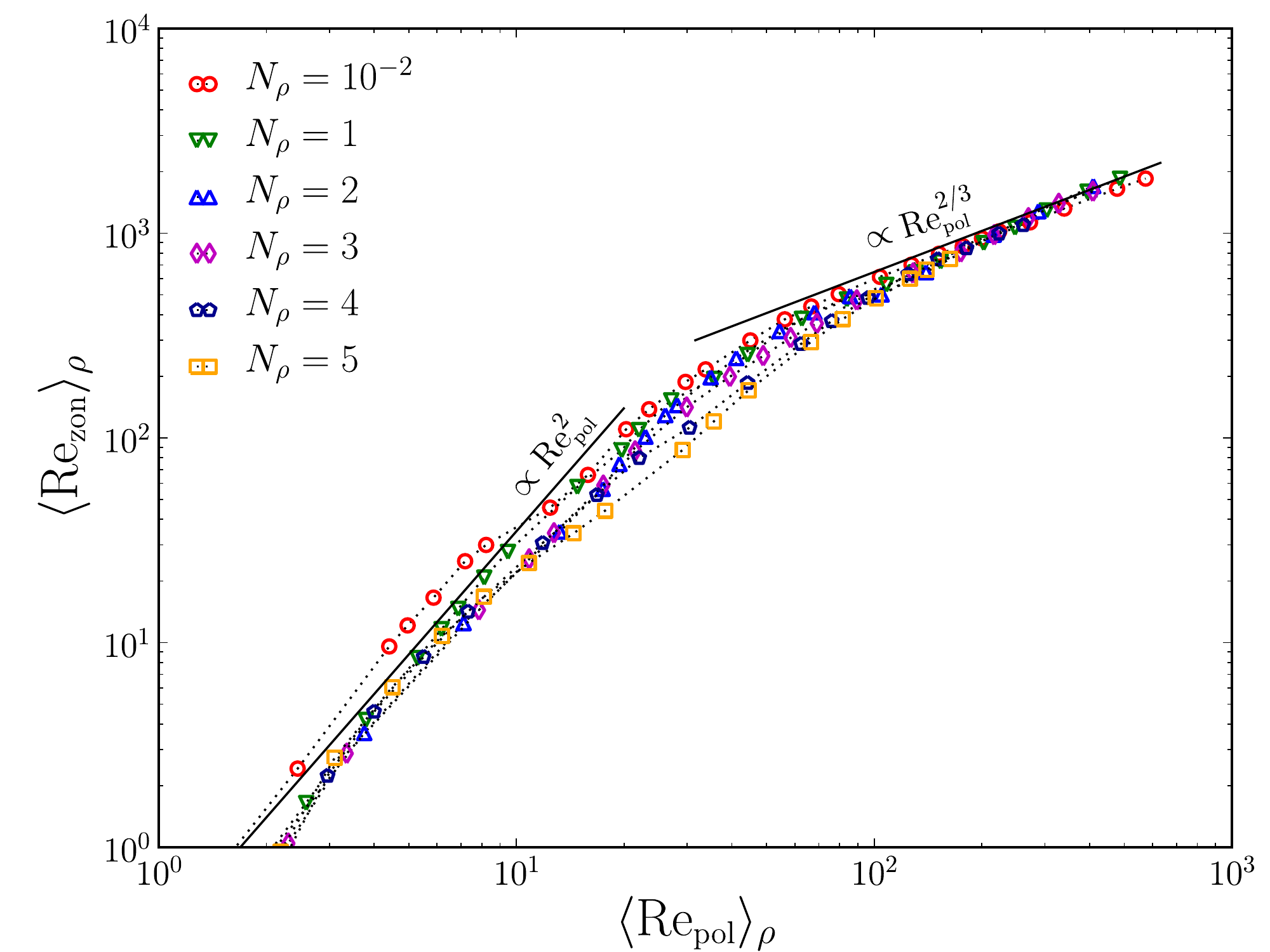}
 \caption{Reynolds number of axisymmetric zonal flows plotted against poloidal
Reynolds number.}
\label{fig:rey}
\end{figure}

\begin{figure}
 \centering
  \includegraphics[width=8cm]{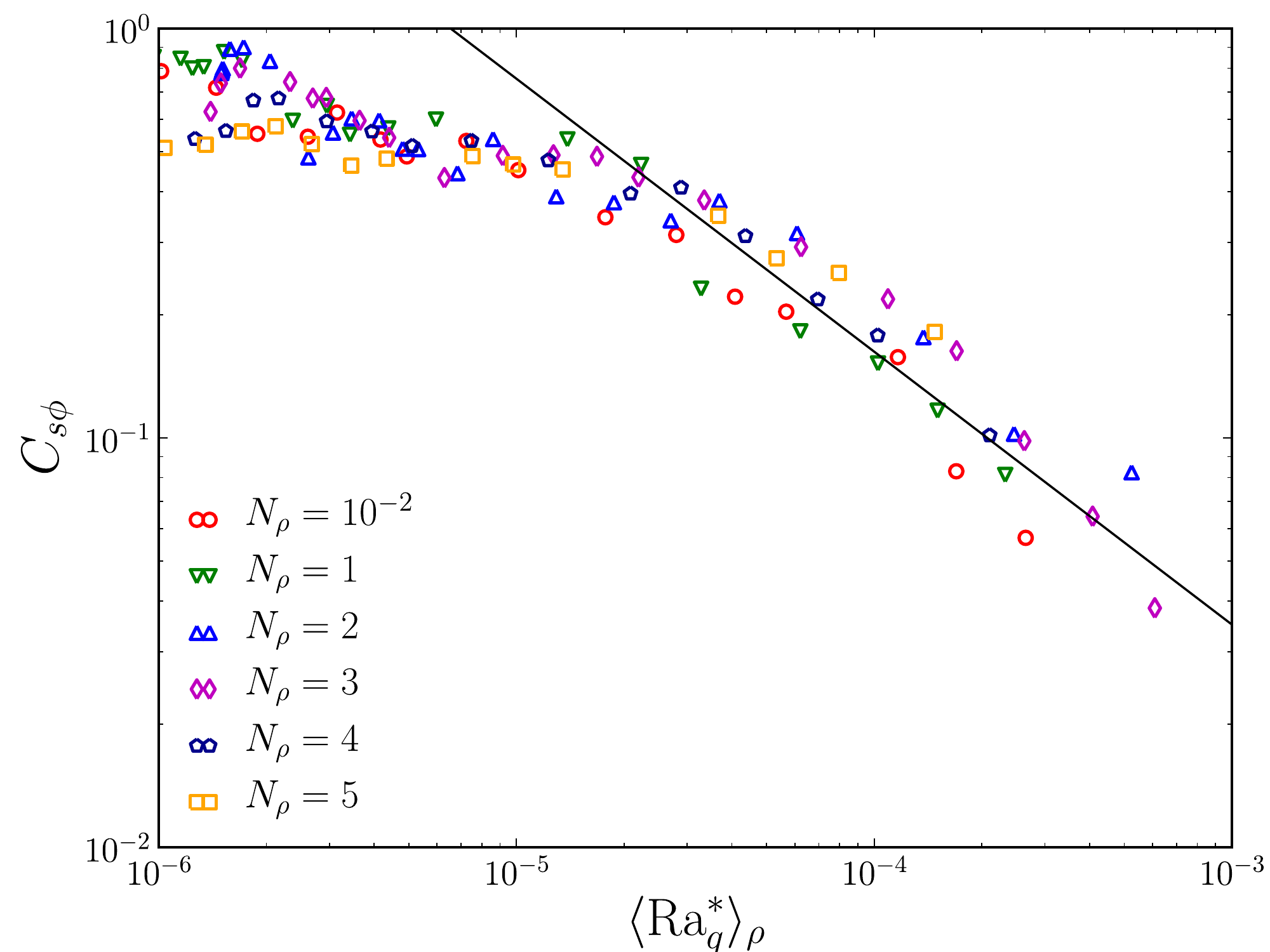}
 \caption{Correlation $C_{s\phi}$ plotted against mass-averaged flux-based
Rayleigh number. The solid line is proportional to $\langle \text{Ra}_q^*
\rangle_\rho^{-2/3}$.}
\label{fig:corr}
\end{figure}

Since the typical length scale of the zonal jets remains of order unity, 
Eq.~(\ref{fig:balance}) suggests the scaling
\citep[e.g.][]{Christensen02}

\begin{equation}
 \langle\text{Re}_{\text{zon}}\rangle_\rho \sim C_{s\phi}\
 \langle\text{Re}_{\text{pol}}\rangle_\rho^2,
 \label{eq:rey_balance}
\end{equation}
where $\langle\text{Re}_{\text{zon}}\rangle_\rho$ is the Reynolds number of
axisymmetric zonal flows and $C_{s\phi}$ quantifies the correlation of $u_s$ and
$u_\phi$ throughout the shell:
  
\begin{equation}
 C_{s\phi} = \dfrac{\left[\ \overline{\rb u_s u_\phi}
\ \right]}{\left[\left(\overline{\rb u_s^2}\ \overline{\rb u_\phi^2}
\right)^{1/2}\right]}\;\;.
\label{eq:correlation}
\end{equation}
Here, the rectangular brackets $[\cdots]$ corresponds to an average over $s$ and $z$.
In Fig.~\ref{fig:rey} we show $\langle\text{Re}_{\text{zon}}\rangle_\rho$
versus  $\langle\text{Re}_{\text{pol}}\rangle_\rho$ for all the simulations of
this study, while Fig.~\ref{fig:corr} shows the dependence of $C_{s\phi}$
on the mass-weighted flux-based Rayleigh number for some snapshots. 

For lower Rayleigh numbers and thus smaller poloidal Reynolds numbers
$\langle \text{Re}_{\text{zon}}\rangle_\rho$ scales roughly like
$\langle\text{Re}_{\text{pol}}\rangle_\rho^2$ and 
$C_{s\phi}$ is largely independent of the Rayleigh number. Convection assumes the 
form of regular large scale columns that are tilted in prograde direction due to the 
height change in the container and the background density stratification
\citep[e.g.][]{Busse83,Busse02}. The integrity of the rolls in z-direction and
the consistent tilt guarantees a good correlation of $u_s$ and $u_\phi$ and
$C_{s\phi}$ is of order one.

For larger Rayleigh numbers the results approach the asymptotic scaling 

\begin{equation}
\label{eqn:rezon2}
\langle\text{Re}_{\text{zon}}\rangle_\rho\sim
\langle\text{Re}_{\text{pol}}\rangle_\rho^{2/3}.
\end{equation}
When using Eq.~(\ref{eq:rossby_scale}) and Eq.~(\ref{eq:rey_balance})
this suggests that 

\begin{equation}
\label{eqn:CRa}
C_{s\phi} \sim \langle \text{Ra}_q^* \rangle_\rho^{-2/3},
\end{equation}
which is more or less confirmed by the correlation shown in
Fig.~\ref{fig:corr}. The small scale turbulent motions that develop for highly
supercritical convection tend to destroy the correlation
\citep{Christensen02,Showman10}.
Once more, neither the ratio of $\langle\text{Re}_{\text{zon}}\rangle_\rho$ to
$\langle\text{Re}_{\text{pol}}\rangle_\rho$, nor the dependence of the
correlation $C_{s\phi}$ on $\langle\text{Ra}_q^* \rangle_\rho$ is significantly
affected by the different density stratifications employed.

The loss of correlation explains why the ratio of total to non-zonal kinetic 
energy shown in Fig.~\ref{fig:resc_ranu}c decreases for larger Rayleigh 
numbers. Our scaling of $C_{s\phi}$ differs from \cite{Showman10}, who suggest 
an exponent of $-1/2$. This also explains the difference in the scaling of the 
Rossby number mentioned above. This difference may come
from specific assumptions used in the model by \citep{Kaspi09}. While these
authors employ a more realistic equation of state, they also consider
a simplified expression of the viscosity and neglect viscous heating.
As \cite{Jones09} have demonstrated, viscous heating can contribute up to
$50\%$ to the heat budget, and is therefore likely to influence the average
flow properties in the strongly nonlinear regime.

\subsection{Density stratification and vorticity production}

The $z$-component of the vorticity equation allows to analyze how 
the correlation between $u_s$ and $u_\phi$ comes about. Taking the
$z$-component of the curl of the momentum equation (\ref{eq:NS}) leads
to

\begin{equation}
\begin{aligned}
 \dfrac{D \omega_z}{D t} -\lp\vec{\omega}\cdot\vec{\nabla}\rp u_z = &
\dfrac{2}{\text{E}}\dfrac{\partial u_z}{\partial z} - \lp \dfrac{2}{\text{E}}
+\omega_z \rp \vec{\nabla}\cdot\vec{u} \\
& -\dfrac{\text{Ra}}{\text{Pr}}\dfrac{r_o^2}{r^3}\dfrac{ \partial s}{\partial
\phi}+\left[\vec{\nabla}\times \lp\rb^{-1}\vec{\nabla}\cdot\tens{S}\rp\right]_z,
\end{aligned}
\label{eq:vortz}
\end{equation}
where $D/Dt$ denotes the substantial time derivative and $\omega_z$ is the
vorticity component along the axis of rotation. For small Ekman numbers, the
first two terms on the right hand side dominate. In the Boussinesq
approximation, the divergence of $\vec{u}$ vanishes and thus only 
the first term can contribute. This is called the vortex stretching term 
because it describes the vorticity changes in a fluid column experiencing the 
height gradient of the container \citep[e.g.][]{Schaeffer05,Glatz09}.
Because of the curved boundary, the absolute value of the gradient and thus the 
stretching effect increases with $s$, leading to the prograde tilt of the 
convective columns \citep[e.g.][]{Busse83,Busse02}.
At larger Rayleigh numbers, these columns loose their integrity along $z$ 
and turbulent effects rather than the boundary curvature start to influence the 
flow. As reported above, this leads to the decrease in the correlation $C_{s\phi}$. 

Besides this classical mechanism, compressibility brings
a new vorticity source. As explained by \cite{Evonuk08} and \cite{Glatz09},
rising (sinking) plumes generate negative (positive) vorticity due to the
compressional source $(2\text{E}^{-1}+\omega_z)\vec{\nabla}\cdot \vec{u}$ in
Eq.~(\ref{eq:vortz}). In the 2-D anelastic simulations performed by
\cite{Glatz09}, the classical vortex stretching term is neglected and
the new compressional term is the only source of vorticity. They report that the
direction of the equatorial jet depends on the variations of the
density background. According to Eq.~(\ref{eq:anel}), this new compressional
source $\vec{\nabla}\cdot\vec{u} = -(d\ln\rb/dr)\ u_r$ is directly proportional
to the density gradient. It is therefore expected to prevail closer to the
outer boundary, where this gradient is strong. Moreover, since this mechanism is
a local process, it may be less sensitive to the loss of correlation along a
convective column with increasing Rayleigh number \citep{Glatz09}.

At first sight, however, this is hard to reconcile with the fact that the
zonal wind structure depends little on the density stratification in our
simulations. Figures \ref{fig:rey} and \ref{fig:corr} have actually demonstrated
that the decorrelation affects all our cases to a similar degree, independently
of the density background.

\begin{figure}
 \centering
  \includegraphics[width=8cm]{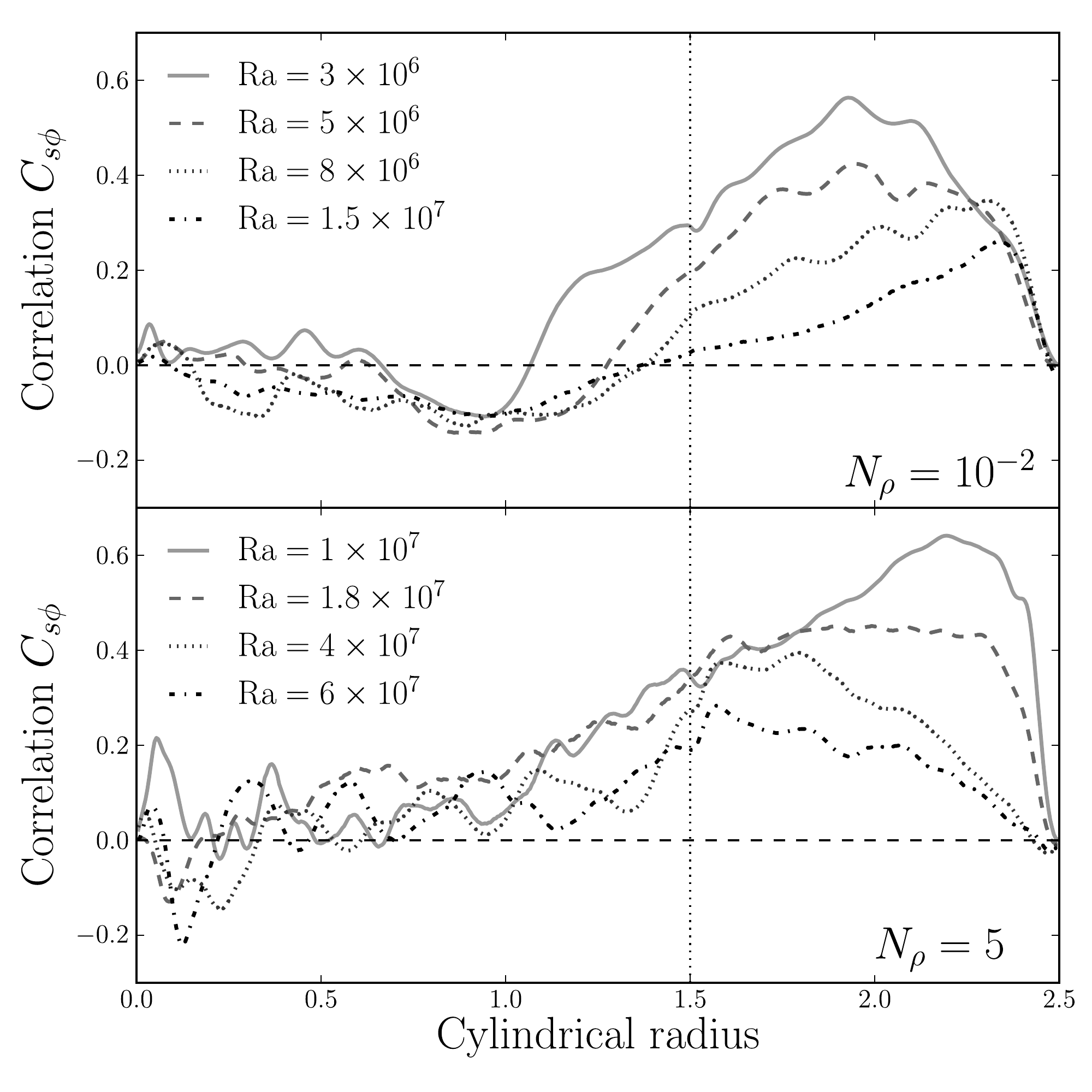}
 \caption{Time-average of the correlation $C_{s\phi}$ integrated over
cylindrical surfaces for simulations with $N_\rho=10^{-2}$ (upper panel):
$\text{Ra}=3\times 10^{6}$ (solid light-grey line), $\text{Ra}=5\times 10^{6}$
(dashed grey line), $\text{Ra}=8\times 10^{6}$ (dotted dark-grey line) and
$\text{Ra}=1.5\times 10^{7}$ (dot-dashed black line); and for simulations with
$N_\rho=5$ (lower panel): $\text{Ra}=1\times 10^{7}$ (solid light-grey line),
$\text{Ra}=1.8\times 10^{7}$ (dashed grey line), $\text{Ra}=4\times 10^{7}$
(dotted dark-grey line) and $\text{Ra}=6\times 10^{7}$ (dot-dashed black line).
The vertical line corresponds to the tangent cylinder.}
\label{fig:corrRa}
\end{figure}

\begin{figure*}
 \centering
  \includegraphics[width=14cm]{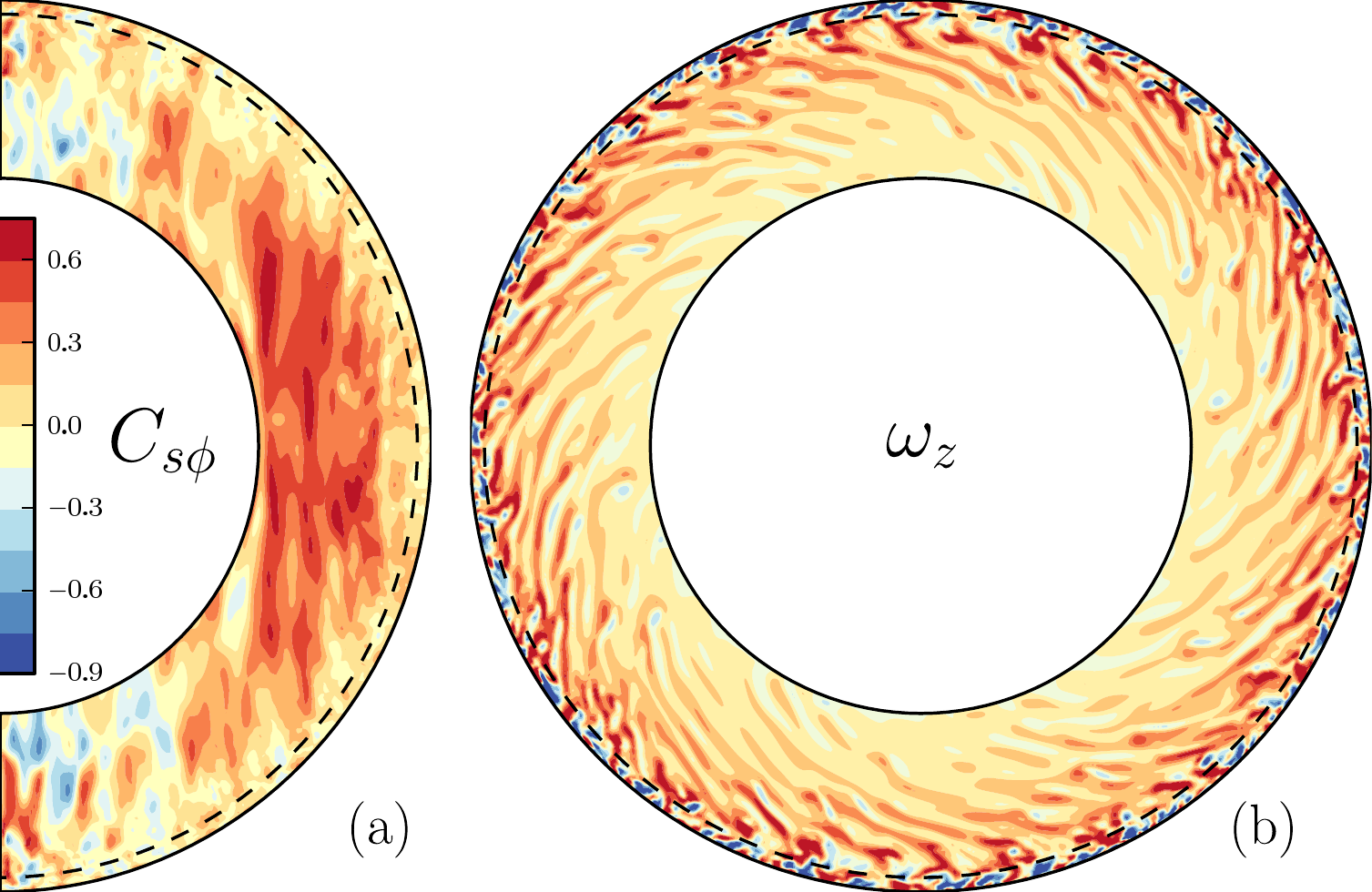}
 \caption{\textbf{a)} Correlation $C_{s\phi}$ in the
meridian plane. \textbf{b)} Vorticity along the axis of rotation $\omega_z$
displayed in the equatorial plane. From a model with $N_\rho=5$,
$\text{Ra}=4\times 10^{7}$. Positive values are rendered in red, negative ones
in blue. The dashed lines in both panels correspond to the radius
where the turnover timescale of convection becomes equal to the timescale
associated with zonal shear (see Eq.~\ref{eq:timecond}).}
\label{fig:reynolds}
\end{figure*}

In order to better understand the decorrelation, Fig.~\ref{fig:corrRa} compares
the radial correlation profiles for nearly Boussinesq and $N_\rho=5$ cases at
various Rayleigh numbers. In the nearly Boussinesq cases, the correlation is
predominantly lost in the interior but changes very little close to the outer
surface, in agreement with the results of \cite{Christensen02}. For the strongly
stratified cases, the correlation indeed peaks close to the outer boundary as
expected from the new compressional vorticity source, as long as the Rayleigh
number remains moderate. However, on increasing the supercriticality, this
particularly strong correlation is lost first and the maximum of $C_{s\phi}$ is
now located deeper in the interior.

\begin{figure}
 \centering
  \includegraphics[width=8cm]{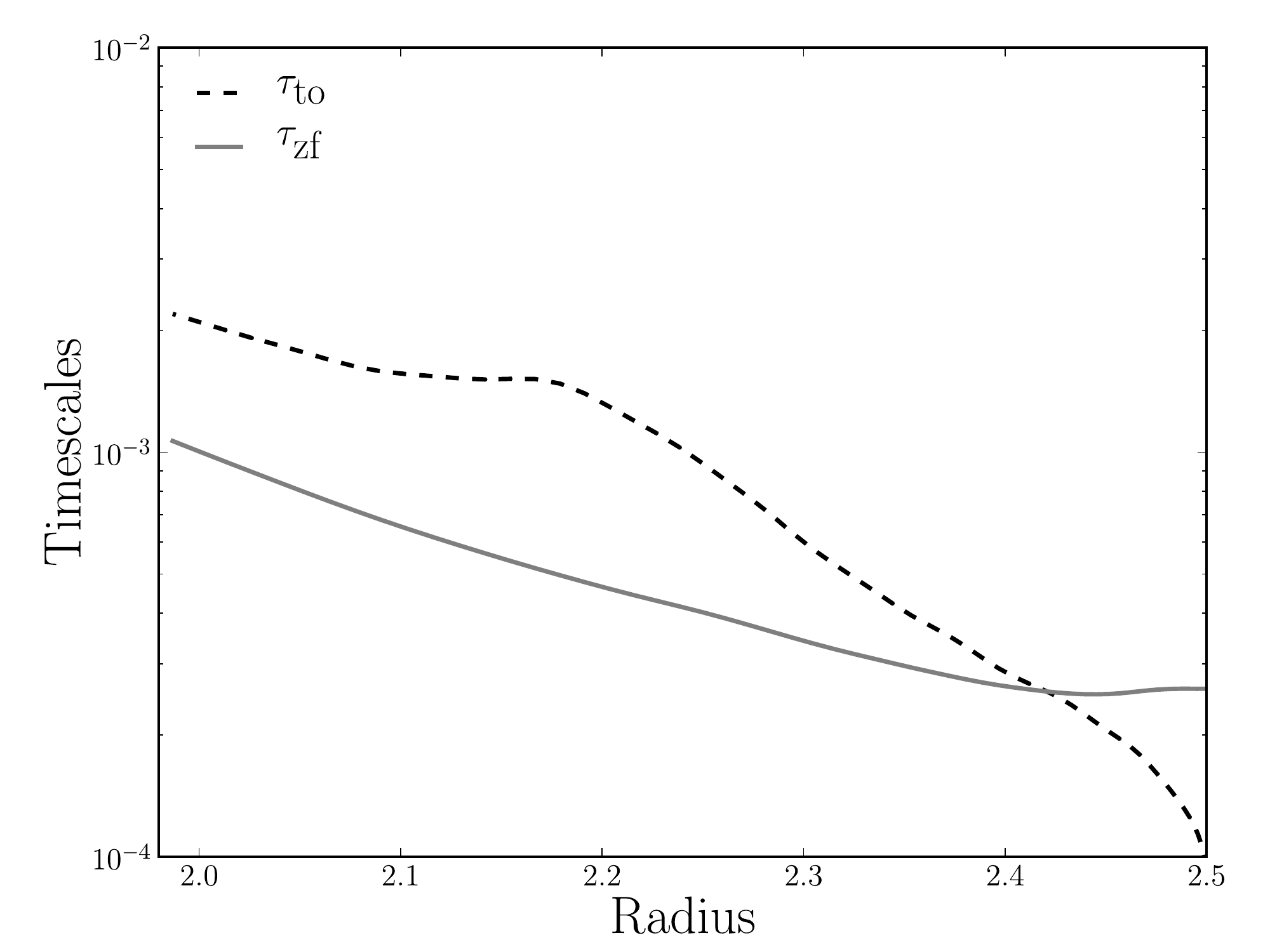}
 \caption{Radial profiles of the two timescales defined in
Eq.~(\ref{eq:timecond}): turnover timescale of convection (dashed black line)
and timescale associated with shear (solid grey line). These profiles have been
derived from an azimuthal average of the values in the equatorial plane for a
model with $N_\rho=5$ and $\text{Ra}=4\times 10^{7}$ (same model as in
Fig.~\ref{fig:reynolds}).}
\label{fig:timescales}
\end{figure}

This is also illustrated by Fig.~\ref{fig:reynolds}a, which shows a meridional
cut of azimuthally averaged $C_{s\phi}$. Figure \ref{fig:reynolds}b shows that
the decorrelation goes along with a change in the vorticity pattern.
In the inner part, the flow still shows a pronounced tilt that leads to Reynolds
stresses and is clearly dominated by positive $z$-vorticity. Close to the outer
boundary, convective motions are more radially oriented, without a preferred
tilt or a clearly preferred sign. We attribute this to an increased mixing
efficiency in the very outer part of the flow \citep{Aurnou07}. The local
turnover timescale of convection is a good proxy to quantify the mixing
efficiency. We estimate this timescale to be $\tau_{\text{to}}\sim
H_\rho/u_{\text{conv}}$. This can be compared with a timescale associated to
zonal shear, defined by $\tau_{\text{zf}} \sim \delta/u_{\text{zonal}}$ (where
$\delta$ is the typical width of the jets). To yield a significant correlation
between $u_s$ and $u_\phi$, the shear timescale needs to be lower than the
turnover timescale, so that a fluid parcel could be diverted in azimuthal
direction before loosing its identity, leading to

\begin{equation}
 \tau_{\text{to}} > \tau_{\text{zf}} \Longrightarrow
\dfrac{H_\rho}{u_{\text{conv}}} >
\dfrac{\delta}{u_{\text{zonal}}}.
\label{eq:timecond}
\end{equation}
Figure~\ref{fig:timescales} shows that this condition is fulfilled in the inner
part of the layer but fails in a very thin layer close to the surface
(approximately the outermost $5\%$). We have marked the radius where both
timescales become equal by a dashed circle in Fig.~\ref{fig:reynolds}.

As demonstrated above for stronger stratifications (see Fig.~\ref{fig:vranel}),
$H_\rho$ decreases outward while $u_{\text{conv}}$ increases. This leads to a
shorter turnover timescale close to the surface.                               
Therefore, surface plumes that could potentially help to generate
vorticity via the compressional source have a too short lifetime to
significantly foster Reynolds stresses.       
In the deeper part, the extra efficiency gained by compressible effects is
limited, explaining why zonal jets of compressible and Boussinesq simulations
are found to be similar.

\section{Conclusion}

\label{sec:conclusion}

We have investigated the effects of compressibility in 3-D
rapidly rotating convection. Following \cite{Glatz1}, \cite{Jones09} and
\cite{Kaspi09}, all the simulations have been computed under the anelastic
approximation. As we focus on the effects of the density stratification, we have
deliberately chosen a moderate Ekman number of $10^{-4}$, which allowed us to
reach more supercritical Rayleigh numbers. From the simulations
computed in this parametric study, we highlight the following results:

\begin{itemize}
\item Concerning the onset of convection, an increase of the density
stratification is accompanied with a strong confinement of the convective
columns close to the outer boundary as well as an increase of the critical
azimuthal wavenumber, in agreement with the results of the linear stability of
\cite{Jones09a}. This gradual outward shift of the onset of convection is
explained by the radial increase of buoyancy.

\item When increasing supercriticality, the solutions which simply drift close
to onset, start to vacillate, then become intermittent and finally chaotic
\citep{Grote01}. While these different regimes are clearly separated in the
Boussinesq cases, the transitions occupy a shorter fraction of the parameter
space with increase of density stratification.

\item For stronger background density stratifications, the convective flow
amplitude increases outward, while the lengthscale decreases. Concerning zonal
winds, Boussinesq and anelastic simulations show very similar properties,
provided the convection is strongly driven: both the extent and the number of
jets become fairly independent of the density background. 
This suggests that zonal flows follow a universal behaviour, independent of the 
density stratification, and may also explain why Boussinesq models were already 
very successful in reproducing the morphology of the jets observed in gas
giants \citep[e.g.][]{Heimpel05}.

\item In the strongly nonlinear regime, the solutions seem to follow the same 
asymptotic scaling laws, independently of the density stratification.
Convective Rossby number and Nusselt number follow  previously studied laws
\citep{Christensen02}. The scaling of the total Rossby number differs slightly
from that suggested by \cite{Showman10}, which may be attributed to the
different model used by these authors. The obtained scaling laws allow us to
extrapolate our simulations to Jupiter. Taking an internal heat flux of
$5.44\text{~W.}\text{m}^{-2}$
\citep{Hanel81}, $\Omega=1.75\times 10^{-4} \text{~s}^{-1}$, 
$c_p=1.3\times 10^{4}\text{~J.kg}^{-1}\text{.K}^{-1}$ and
$d=1.2\times 10^{4}\text{~kms}$, the evolution models of \cite{Guillot04}
suggest that the flux-based Rayleigh number decreases from
$\text{Ra}_q^* \sim 5\times 10^{-7}$ at the surface to $\text{Ra}_q^*
\sim3\times  10^{-12}$ at the bottom of the molecular envelope (i.e.  
$\eta=0.85$).
This leads to a mass-weighted flux based Rayleigh number of
$\langle\text{Ra}_q^*\rangle_\rho\sim 1.2\times 10^{-11}$. 
Our scaling law (Eq.~\ref{eq:rossby_scale}) then predicts a Rossby number of
$6\times10^{-4}$, which corresponds to an average zonal velocity around
$1\text{~m.}\text{s}^{-1}$. In agreement with the previous extrapolations of
\cite{Showman10}, this value is much weaker (about one order of magnitude) than
the surface zonal winds of gas giants. This rescaling is however tentative as
the scaling laws may still depend on other parameters that have been kept
constant in this study, for example Ekman number, Prandtl number or aspect
ratio, and further simulations would be required to clarify this.

\item  The zonal jets are maintained by Reynolds stresses, which rely on the
correlation between zonal ($u_\phi$) and cylindrically radial ($u_s$) flow
components. The gradual loss of this correlation with increasing
supercriticality hampers all our simulations, independently of the background
density stratification. However in the strongly stratified simulations, the
correlation is first lost in the outer part of the flow, where compressibility
effects are most significant. Because of the short lengthscales and the large
flow amplitudes found there, the fluid parcels loose their identity before
interacting efficiently with the zonal flows. This may explain why the
additional compressional source of vorticity has not the important effect
expected by \cite{Evonuk08} and \cite{Glatz09} in our simulations. 
However, the influence of the different vorticity contributions are hard to
disentangle in 3-D compressible convection, stressing the need of
future theoretical work to address this question.

\end{itemize}

All the results obtained in this study have been derived in a non-conducting
layer, where magnetic effects have been neglected. Nevertheless, as suggested by
\cite{Liu08}, the Ohmic dissipation associated with the outer semi-conducting
region could strongly limit the extent of the surface jets. Addressing this
problem would require 3-D dynamo simulations of compressible convection with
variable conductivity \citep{Stanley09,Heimpel11}.

\section*{Acknowledgements}

All the computations have been carried out on the GWDG computer facilities in
G\"ottingen. This work was supported by the Special Priority Program 1488
(PlanetMag, \url{http://www.planetmag.de}) of the German Science Foundation.

\bibliographystyle{model2-names}

\begin{table*}
\centering
 \caption{List of numerical simulations presented in this work. For all the
simulations, $\text{E}=10^{-4}$ and $\text{Pr}=1$.}
{\scriptsize
\begin{tabular}{ccccc}
\toprule
${Ra}/{Ra}_{\text{crit}}$ & $\langle \text{Ro}\rangle_\rho$ & $\langle
\text{Ro}_\text{pol}\rangle_\rho$ & $E_{\text{tor}}^{m=0}/E_{\text{kin}}$ &
Nu \\
\midrule
\multicolumn{5}{c}{$N_\rho=10^{-2}$, symbol=red circle} \\
\midrule
$ 1.0006 $ & $ 4.49 \times 10^{-5} $ & $ 2.49 \times 10^{-5} $ & $ 0.002 $ & $
1.0002 $ \\
$ 1.0063 $ & $ 1.03 \times 10^{-4} $ & $ 5.71 \times 10^{-5} $ & $ 0.011 $ & $
1.0009 $ \\
$ 1.035 $ & $ 2.35 \times 10^{-4} $ & $ 1.26 \times 10^{-4} $ & $ 0.058 $ & $
1.0045 $ \\
$ 1.15 $ & $ 5.04 \times 10^{-4} $ & $ 2.45 \times 10^{-4} $ & $ 0.231 $ & $
1.016 $ \\
$ 1.72 $ & $ 1.25 \times 10^{-3} $ & $ 4.41 \times 10^{-4} $ & $ 0.586 $ & $
1.045 $ \\
$ 2.01 $ & $ 1.52 \times 10^{-3} $ & $ 4.96 \times 10^{-4} $ & $ 0.640 $ & $
1.053 $ \\
$ 2.30 $ & $ 1.98 \times 10^{-3} $ & $ 5.86 \times 10^{-4} $ & $ 0.700 $ & $
1.067 $ \\
$ 2.87 $ & $ 2.85 \times 10^{-3} $ & $ 7.19 \times 10^{-4} $ & $ 0.773 $ & $
1.087 $ \\
$ 3.45 $ & $ 3.37 \times 10^{-3} $ & $ 8.23 \times 10^{-4} $ & $ 0.788 $ & $
1.104 $ \\
$ 4.60 $ & $ 5.05 \times 10^{-3} $ & $ 1.24 \times 10^{-3} $ & $ 0.816 $ & $
1.1800 $ \\
$ 5.75$ & $ 7.18 \times 10^{-3} $ & $ 1.58 \times 10^{-3} $ & $ 0.843 $ & $
1.23 $ \\
$ 7.47 $ & $ 1.16 \times 10^{-2} $ & $ 2.03 \times 10^{-3} $ & $ 0.902 $ & $
1.31 $ \\
$ 8.62 $ & $ 1.44 \times 10^{-2} $ & $ 2.35 \times 10^{-3} $ & $ 0.916 $ & $
1.37 $ \\
$ 10.35 $ & $ 1.95 \times 10^{-2} $ & $ 2.97 \times 10^{-3} $ & $ 0.929 $ & $
1.51 $ \\
$ 11.50 $ & $ 2.24 \times 10^{-2} $ & $ 3.38 \times 10^{-3} $ & $ 0.933 $ & $
1.61 $ \\
$ 14.37 $ & $ 3.09 \times 10^{-2} $ & $ 4.51 \times 10^{-3} $ & $ 0.942 $ & $
1.89 $ \\
$ 17.25 $ & $ 3.90 \times 10^{-2} $ & $ 5.62 \times 10^{-3} $ & $ 0.945 $ & $
2.20 $ \\
$ 20.12 $ & $ 4.51 \times 10^{-2} $ & $ 6.67 \times 10^{-3} $ & $ 0.943 $ & $
2.52 $ \\
$ 23.00 $ & $ 5.20 \times 10^{-2} $ & $ 7.96 \times 10^{-3} $ & $ 0.939 $ & $
2.89 $ \\
$ 28.75 $ & $ 6.34 \times 10^{-2} $ & $ 1.04 \times 10^{-2} $ & $ 0.932 $ & $
3.65 $ \\
$ 34.50 $ & $ 7.30 \times 10^{-2} $ & $ 1.27 \times 10^{-2} $ & $ 0.924 $ & $
4.44 $ \\
$ 40.25 $ & $ 8.30 \times 10^{-2} $ & $ 1.52 \times 10^{-2} $ & $ 0.915 $ & $
5.29 $ \\
$ 46.00 $ & $ 9.10 \times 10^{-2} $ & $ 1.77 \times 10^{-2} $ & $ 0.903 $ & $
6.04 $ \\
$ 51.75 $ & $ 9.98 \times 10^{-2} $ & $ 2.00 \times 10^{-2} $ & $ 0.893 $ & $
6.81 $ \\
$ 57.50 $ & $ 1.09 \times 10^{-1} $ & $ 2.24 \times 10^{-2} $ & $ 0.883 $ & $
7.60 $ \\
$ 69.00 $ & $ 1.21 \times 10^{-1} $ & $ 2.73 \times 10^{-2} $ & $ 0.874 $ & $
9.22 $ \\
$ 86.25 $ & $ 1.42 \times 10^{-1} $ & $ 3.40 \times 10^{-2} $ & $ 0.863 $ & $
11.53 $ \\
$ 115.0 $ & $ 1.79 \times 10^{-1} $ & $ 4.78 \times 10^{-2} $ & $ 0.845 $ & $
16.21 $ \\
$ 143.7 $ & $ 2.03 \times 10^{-1} $ & $ 5.74 \times 10^{-2} $ & $ 0.830 $ & $
19.50 $ \\
\midrule
\multicolumn{5}{c}{$N_\rho=2$, symbol=blue triangle} \\
\midrule
$ 1.0001 $ & $ 3.65 \times 10^{-5} $ & $ 2.63 \times 10^{-5} $ & $ 0.001 $ & $
1.0001 $ \\
$ 1.0035 $ & $ 9.13 \times 10^{-5} $ & $ 6.77 \times 10^{-5} $ & $ 0.007 $ & $
1.0005 $ \\
$ 1.0079 $ & $ 1.28 \times 10^{-4} $ & $ 9.57 \times 10^{-5} $ & $ 0.014 $ & $
1.0010 $ \\
$ 1.051 $ & $ 2.80 \times 10^{-4} $ & $ 1.99 \times 10^{-4} $ & $ 0.093 $ & $
1.0045 $ \\
$ 1.13 $ & $ 5.96 \times 10^{-4} $ & $ 3.75 \times 10^{-4} $ & $ 0.339 $ & $
1.013 $ \\
$ 1.31 $ & $ 1.56 \times 10^{-3} $ & $ 7.12 \times 10^{-4} $ & $ 0.636 $ & $
1.040 $ \\
$ 1.57 $ & $ 3.94 \times 10^{-3} $ & $ 1.32 \times 10^{-3} $ & $ 0.782 $ & $
1.11 $ \\
$ 1.75 $ & $ 6.15 \times 10^{-3} $ & $ 1.74 \times 10^{-3} $ & $ 0.838 $ & $
1.17 $ \\
$ 1.92 $ & $ 7.95 \times 10^{-3} $ & $ 1.94 \times 10^{-3} $ & $ 0.877 $ & $
1.20 $ \\
$ 2.19 $ & $ 1.07 \times 10^{-2} $ & $ 2.29 \times 10^{-3} $ & $ 0.903 $ & $
1.26 $ \\
$ 2.45 $ & $ 1.35 \times 10^{-2} $ & $ 2.61 \times 10^{-3} $ & $ 0.920 $ & $
1.31 $ \\
$ 2.62 $ & $ 1.50 \times 10^{-2} $ & $ 2.81 \times 10^{-3} $ & $ 0.925 $ & $
1.35 $ \\
$ 3.06 $ & $ 2.04 \times 10^{-2} $ & $ 3.49 \times 10^{-3} $ & $ 0.936 $ & $
1.49 $ \\
$ 3.50 $ & $ 2.53 \times 10^{-2} $ & $ 4.11 \times 10^{-3} $ & $ 0.940 $ & $
1.64 $ \\
$ 4.38 $ & $ 3.43 \times 10^{-2} $ & $ 5.44 \times 10^{-3} $ & $ 0.941 $ & $
1.97 $ \\
$ 5.25 $ & $ 4.24 \times 10^{-2} $ & $ 6.81 \times 10^{-3} $ & $ 0.938 $ & $
2.38 $ \\
$ 6.13 $ & $ 5.10 \times 10^{-2} $ & $ 8.57 \times 10^{-3} $ & $ 0.931 $ & $
2.94 $ \\
$ 7.01 $ & $ 5.27 \times 10^{-2} $ & $ 1.05 \times 10^{-2} $ & $ 0.905 $ & $
3.52 $ \\
$ 8.76 $ & $ 6.80 \times 10^{-2} $ & $ 1.40 \times 10^{-2} $ & $ 0.900 $ & $
4.64 $ \\
$ 13.14 $ & $ 1.03 \times 10^{-1} $ & $ 2.16 \times 10^{-2} $ & $ 0.904 $ & $
6.98 $ \\
$ 17.52 $ & $ 1.35 \times 10^{-1} $ & $ 2.88 \times 10^{-2} $ & $ 0.902 $ & $
9.38 $ \\
$ 26.29 $ & $ 1.80 \times 10^{-1} $ & $ 4.10 \times 10^{-2} $ & $ 0.896 $ & $
13.33 $ \\
\midrule
\multicolumn{5}{c}{$N_\rho=4$, symbol=dark blue pentagon} \\
\midrule
$ 1.0005 $ & $ 2.52 \times 10^{-5} $ & $ 2.02 \times 10^{-5} $ & $ 0.000 $ & $
1.0002 $ \\
$ 1.0016 $ & $ 3.63 \times 10^{-5} $ & $ 3.02 \times 10^{-5} $ & $ 0.000 $ & $
1.0005 $ \\
$ 1.0097 $ & $ 8.11 \times 10^{-5} $ & $ 6.89 \times 10^{-5} $ & $ 0.002 $ & $
1.0025 $ \\
$ 1.025 $ & $ 1.34 \times 10^{-4} $ & $ 1.11 \times 10^{-4} $ & $ 0.006 $ & $
1.0063 $ \\
$ 1.079 $ & $ 2.21 \times 10^{-4} $ & $ 1.82 \times 10^{-4} $ & $ 0.039 $ & $
1.017 $ \\
$ 1.34 $ & $ 4.11 \times 10^{-4} $ & $ 2.96 \times 10^{-4} $ & $ 0.280 $ & $
1.037 $ \\
$ 1.61 $ & $ 6.76 \times 10^{-4} $ & $ 4.00 \times 10^{-4} $ & $ 0.462 $ & $
1.054 $ \\
$ 1.88 $ & $ 1.09 \times 10^{-3} $ & $ 5.50 \times 10^{-4} $ & $ 0.607 $ & $
1.078 $ \\
$ 2.15 $ & $ 1.69 \times 10^{-3} $ & $ 7.32 \times 10^{-4} $ & $ 0.706 $ & $
1.109 $ \\
$ 2.69 $ & $ 3.41 \times 10^{-3} $ & $ 1.18 \times 10^{-3} $ & $ 0.809 $ & $
1.21 $ \\
$ 3.23 $ & $ 5.70 \times 10^{-3} $ & $ 1.68 \times 10^{-3} $ & $ 0.854 $ & $
1.35 $ \\
$ 3.77 $ & $ 8.52 \times 10^{-3} $ & $ 2.20 \times 10^{-3} $ & $ 0.877 $ & $
1.50 $ \\
$ 4.31 $ & $ 1.20 \times 10^{-2} $ & $ 3.05 \times 10^{-3} $ & $ 0.869 $ & $
1.92 $ \\
$ 5.39 $ & $ 1.96 \times 10^{-2} $ & $ 4.43 \times 10^{-3} $ & $ 0.893 $ & $
2.52 $ \\
$ 7.01 $ & $ 3.03 \times 10^{-2} $ & $ 6.25 \times 10^{-3} $ & $ 0.907 $ & $
3.29 $ \\
$ 8.63 $ & $ 3.89 \times 10^{-2} $ & $ 7.61 \times 10^{-3} $ & $ 0.916 $ & $
3.70 $ \\
$ 10.79 $ & $ 5.04 \times 10^{-2} $ & $ 9.62 \times 10^{-3} $ & $ 0.919 $ & $
4.48 $ \\
$ 14.03 $ & $ 6.57 \times 10^{-2} $ & $ 1.25 \times 10^{-2} $ & $ 0.920 $ & $
5.49 $ \\
$ 17.27 $ & $ 7.80 \times 10^{-2} $ & $ 1.51 \times 10^{-2} $ & $ 0.917 $ & $
6.56 $ \\
$ 21.59 $ & $ 8.86 \times 10^{-2} $ & $ 1.81 \times 10^{-2} $ & $ 0.909 $ & $
7.87 $ \\
$ 26.99 $ & $ 1.05 \times 10^{-1} $ & $ 2.23 \times 10^{-2} $ & $ 0.904 $ & $
8.65 $ \\
$ 32.39 $ & $ 1.16 \times 10^{-1} $ & $ 2.61 \times 10^{-2} $ & $ 0.896 $ & $
9.73 $ \\
\bottomrule
 \end{tabular}
 \hspace{1cm}
 \begin{tabular}{ccccc}
\toprule
${Ra}/{Ra}_{\text{crit}}$ & $\langle \text{Ro}\rangle_\rho$ & $\langle
\text{Ro}_\text{pol}\rangle_\rho$ & $E_{\text{tor}}^{m=0}/E_{\text{kin}}$ &
Nu \\
\midrule
\multicolumn{5}{c}{$N_\rho=1$, symbol=green triangle} \\
\midrule
$ 1.0010 $ & $ 5.36 \times 10^{-5} $ & $ 3.66 \times 10^{-5} $ & $ 0.002 $ & $
1.0001 $ \\
$ 1.0068 $ & $ 1.31 \times 10^{-4} $ & $ 8.82 \times 10^{-5} $ & $ 0.014 $ & $
1.0008 $ \\
$ 1.0145 $ & $ 1.95 \times 10^{-4} $ & $ 1.30 \times 10^{-4} $ & $ 0.033 $ & $
1.0018 $ \\
$ 1.024 $ & $ 2.54 \times 10^{-4} $ & $ 1.67 \times 10^{-4} $ & $ 0.055 $ & $
1.0029 $ \\
$ 1.062 $ & $ 4.23 \times 10^{-4} $ & $ 2.58 \times 10^{-4} $ & $ 0.156 $ & $
1.007 $ \\
$ 1.15 $ & $ 6.99 \times 10^{-4} $ & $ 3.79 \times 10^{-4} $ & $ 0.357 $ & $
1.009 $ \\
$ 1.35 $ & $ 1.15 \times 10^{-3} $ & $ 5.29 \times 10^{-4} $ & $ 0.537 $ & $
1.028 $ \\
$ 1.44 $ & $ 1.48 \times 10^{-3} $ & $ 6.17 \times 10^{-4} $ & $ 0.619 $ & $
1.036 $ \\
$ 1.54 $ & $ 1.78 \times 10^{-3} $ & $ 6.88 \times 10^{-4} $ & $ 0.674 $ & $
1.044 $ \\
$ 1.73 $ & $ 2.40 \times 10^{-3} $ & $ 8.13 \times 10^{-4} $ & $ 0.748 $ & $
1.058 $ \\
$ 1.93 $ & $ 3.11 \times 10^{-3} $ & $ 9.47 \times 10^{-4} $ & $ 0.795 $ & $
1.07 $ \\
$ 2.51 $ & $ 6.21 \times 10^{-3} $ & $ 1.48 \times 10^{-3} $ & $ 0.858 $ & $
1.14 $ \\
$ 2.89 $ & $ 9.23 \times 10^{-3} $ & $ 1.97 \times 10^{-3} $ & $ 0.885 $ & $
1.23 $ \\
$ 3.28 $ & $ 1.15 \times 10^{-2} $ & $ 2.20 \times 10^{-3} $ & $ 0.905 $ & $
1.26 $ \\
$ 3.86 $ & $ 1.59 \times 10^{-2} $ & $ 2.71 \times 10^{-3} $ & $ 0.925 $ & $
1.37 $ \\
$ 4.83 $ & $ 2.00 \times 10^{-2} $ & $ 3.57 \times 10^{-3} $ & $ 0.922 $ & $
1.49 $ \\
$ 5.79 $ & $ 2.61 \times 10^{-2} $ & $ 4.43 \times 10^{-3} $ & $ 0.931 $ & $
1.69 $ \\
$ 7.72 $ & $ 3.92 \times 10^{-2} $ & $ 6.28 \times 10^{-3} $ & $ 0.937 $ & $
2.17 $ \\
$ 9.66 $ & $ 4.91 \times 10^{-2} $ & $ 8.39 \times 10^{-3} $ & $ 0.927 $ & $
2.79 $ \\
$ 11.59 $ & $ 5.83 \times 10^{-2} $ & $ 1.08 \times 10^{-2} $ & $ 0.916 $ & $
3.42 $ \\
$ 13.52 $ & $ 6.74 \times 10^{-2} $ & $ 1.31 \times 10^{-2} $ & $ 0.909 $ & $
4.14 $ \\
$ 15.45 $ & $ 7.63 \times 10^{-2} $ & $ 1.53 \times 10^{-2} $ & $ 0.904 $ & $
4.86 $ \\
$ 19.32 $ & $ 9.45 \times 10^{-2} $ & $ 2.02 \times 10^{-2} $ & $ 0.896 $ & $
6.41 $ \\
$ 23.18 $ & $ 1.12 \times 10^{-1} $ & $ 2.48 \times 10^{-2} $ & $ 0.893 $ & $
7.84 $ \\
$ 28.98 $ & $ 1.38 \times 10^{-1} $ & $ 3.04 \times 10^{-2} $ & $ 0.875 $ & $
9.69 $ \\
$ 38.64 $ & $ 1.71 \times 10^{-1} $ & $ 3.95 \times 10^{-2} $ & $ 0.873 $ & $
12.86 $ \\
$ 48.30 $ & $ 1.98 \times 10^{-1} $ & $ 4.87 \times 10^{-2} $ & $ 0.871 $ & $
15.97 $ \\
\midrule
\multicolumn{5}{c}{$N_\rho=3$, symbol=magenta diamond} \\
\midrule
$ 1.0020 $ & $ 4.24 \times 10^{-5} $ & $ 3.32 \times 10^{-5} $ & $ 0.001 $ & $
1.0003 $ \\
$ 1.0039 $ & $ 6.35 \times 10^{-5} $ & $ 5.15 \times 10^{-5} $ & $ 0.002 $ & $
1.0007 $ \\
$ 1.0072 $ & $ 9.32 \times 10^{-5} $ & $ 7.21 \times 10^{-5} $ & $ 0.004 $ & $
1.0013 $ \\
$ 1.013 $ & $ 1.27 \times 10^{-4} $ & $ 1.01 \times 10^{-4} $ & $ 0.008 $ & $
1.0025 $ \\
$ 1.046 $ & $ 2.29 \times 10^{-4} $ & $ 1.75 \times 10^{-4} $ & $ 0.037 $ & $
1.0076 $ \\
$ 1.11 $ & $ 3.14 \times 10^{-4} $ & $ 2.30 \times 10^{-4} $ & $ 0.111 $ & $
1.0134 $ \\
$ 1.24 $ & $ 5.17 \times 10^{-4} $ & $ 3.35 \times 10^{-4} $ & $ 0.307 $ & $
1.024 $ \\
$ 1.63 $ & $ 1.75 \times 10^{-3} $ & $ 7.85 \times 10^{-4} $ & $ 0.675 $ & $
1.074 $ \\
$ 1.83 $ & $ 2.91 \times 10^{-3} $ & $ 1.09 \times 10^{-3} $ & $ 0.770 $ & $
1.11 $ \\
$ 1.96 $ & $ 3.80 \times 10^{-3} $ & $ 1.27 \times 10^{-3} $ & $ 0.812 $ & $
1.13 $ \\
$ 2.28 $ & $ 6.30 \times 10^{-3} $ & $ 1.75 \times 10^{-3} $ & $ 0.866 $ & $
1.20 $ \\
$ 2.61 $ & $ 9.09 \times 10^{-3} $ & $ 2.15 \times 10^{-3} $ & $ 0.899 $ & $
1.27 $ \\
$ 3.27 $ & $ 1.47 \times 10^{-2} $ & $ 2.99 \times 10^{-3} $ & $ 0.920 $ & $
1.45 $ \\
$ 3.92 $ & $ 2.07 \times 10^{-2} $ & $ 3.95 \times 10^{-3} $ & $ 0.924 $ & $
1.76 $ \\
$ 4.57 $ & $ 2.62 \times 10^{-2} $ & $ 4.90 \times 10^{-3} $ & $ 0.922 $ & $
2.09 $ \\
$ 5.23 $ & $ 3.21 \times 10^{-2} $ & $ 5.84 \times 10^{-3} $ & $ 0.924 $ & $
2.42 $ \\
$ 5.88 $ & $ 3.77 \times 10^{-2} $ & $ 6.91 \times 10^{-3} $ & $ 0.922 $ & $
2.81 $ \\
$ 7.19 $ & $ 4.91 \times 10^{-2} $ & $ 8.94 \times 10^{-3} $ & $ 0.922 $ & $
3.51 $ \\
$ 9.81 $ & $ 6.66 \times 10^{-2} $ & $ 1.28 \times 10^{-2} $ & $ 0.916 $ & $
4.80 $ \\
$ 13.08 $ & $ 8.48 \times 10^{-2} $ & $ 1.74 \times 10^{-2} $ & $ 0.908 $ & $
6.30 $ \\
$ 16.35 $ & $ 1.03 \times 10^{-1} $ & $ 2.17 \times 10^{-2} $ & $ 0.905 $ & $
7.84 $ \\
$ 20.92 $ & $ 1.25 \times 10^{-1} $ & $ 2.70 \times 10^{-2} $ & $ 0.902 $ & $
9.47 $ \\
$ 26.16 $ & $ 1.47 \times 10^{-1} $ & $ 3.28 \times 10^{-2} $ & $ 0.900 $ & $
11.76 $ \\
$ 32.70 $ & $ 1.70 \times 10^{-1} $ & $ 4.09 \times 10^{-2} $ & $ 0.886 $ & $
14.14 $ \\
\midrule
\multicolumn{5}{c}{$N_\rho=5$, symbol=orange square} \\
\midrule
$ 1.0017 $ & $ 2.49 \times 10^{-5} $ & $ 2.24 \times 10^{-5} $ & $ 0.000 $ & $
1.0005 $ \\
$ 1.0038 $ & $ 3.56 \times 10^{-5} $ & $ 3.21 \times 10^{-5} $ & $ 0.001 $ & $
1.0011 $ \\
$ 1.025 $ & $ 9.09 \times 10^{-5} $ & $ 8.06 \times 10^{-5} $ & $ 0.003 $ & $
1.0066 $ \\
$ 1.067 $ & $ 1.50 \times 10^{-4} $ & $ 1.33 \times 10^{-4} $ & $ 0.013 $ & $
1.017 $ \\
$ 1.28 $ & $ 2.71 \times 10^{-4} $ & $ 2.19 \times 10^{-4} $ & $ 0.124 $ & $
1.041 $ \\
$ 1.70 $ & $ 4.61 \times 10^{-4} $ & $ 3.10 \times 10^{-4} $ & $ 0.351 $ & $
1.061 $ \\
$ 2.13 $ & $ 8.20 \times 10^{-4} $ & $ 4.50 \times 10^{-4} $ & $ 0.548 $ & $
1.104 $ \\
$ 2.56 $ & $ 1.32 \times 10^{-3} $ & $ 6.19 \times 10^{-4} $ & $ 0.661 $ & $
1.163 $ \\
$ 2.99 $ & $ 1.96 \times 10^{-3} $ & $ 8.10 \times 10^{-4} $ & $ 0.726 $ & $
1.24 $ \\
$ 3.41 $ & $ 2.82 \times 10^{-3} $ & $ 1.08 \times 10^{-3} $ & $ 0.751 $ & $
1.37 $ \\
$ 3.84 $ & $ 3.92 \times 10^{-3} $ & $ 1.44 \times 10^{-3} $ & $ 0.759 $ & $
1.57 $ \\
$ 4.27 $ & $ 5.04 \times 10^{-3} $ & $ 1.77 \times 10^{-3} $ & $ 0.766 $ & $
1.77 $ \\
$ 5.55 $ & $ 9.65 \times 10^{-3} $ & $ 2.92 \times 10^{-3} $ & $ 0.821 $ & $
2.37 $ \\
$ 6.40 $ & $ 1.30 \times 10^{-2} $ & $ 3.56 \times 10^{-3} $ & $ 0.853 $ & $
2.67 $ \\
$ 7.68 $ & $ 1.82 \times 10^{-2} $ & $ 4.45 \times 10^{-3} $ & $ 0.881 $ & $
3.06 $ \\
$ 11.11 $ & $ 3.09 \times 10^{-2} $ & $ 6.65 \times 10^{-3} $ & $ 0.904 $ & $
4.01 $ \\
$ 13.67 $ & $ 3.99 \times 10^{-2} $ & $ 8.17 \times 10^{-3} $ & $ 0.911 $ & $
4.68 $ \\
$ 17.09 $ & $ 5.03 \times 10^{-2} $ & $ 1.01 \times 10^{-2} $ & $ 0.914 $ & $
5.45 $ \\
$ 21.36 $ & $ 6.26 \times 10^{-2} $ & $ 1.26 \times 10^{-2} $ & $ 0.914 $ & $
6.52 $ \\
$ 25.63 $ & $ 6.94 \times 10^{-2} $ & $ 1.40 \times 10^{-2} $ & $ 0.915 $ & $
6.87 $ \\
$ 32.04 $ & $ 7.85 \times 10^{-2} $ & $ 1.63 \times 10^{-2} $ & $ 0.911 $ & $
8.04 $ \\
\bottomrule
 \end{tabular}}
\label{tab:simus}
\end{table*}


\end{document}